\documentclass[11pt]{article}
\usepackage{amsmath,color}
\usepackage{amssymb}
\usepackage{amsfonts}
\usepackage{graphicx}
\usepackage{color}

\definecolor{darkgreen}{rgb}{0,0.35,0}

\numberwithin{equation}{section}

\begin{document}

\title{\textbf{One-loop amplitudes of winding strings in AdS$_3$ and the Coulomb gas approach}}
\author{Gaston Giribet}
\date{}
\maketitle

\begin{center}

\smallskip
\smallskip
\centerline{$^1$ Universit\'{e} Libre de Bruxelles and International Solvay Institutes}
\centerline{{\it ULB-Campus Plaine CPO231, B-1050 Brussels, Belgium.}}

\medskip
\textit{$^{2}$}Departamento de F\'{\i}sica, Universidad de Buenos Aires
FCEN-UBA and IFIBA-CONICET,

\textit{Ciudad Universitaria, Pabell\'{o}n I, 1428, Buenos Aires, Argentina.}



\end{center}

\bigskip

\bigskip

\bigskip

\bigskip

\bigskip

\bigskip

\bigskip

\begin{abstract}

We discuss a Coulomb gas realization of $n$-point correlation functions in the $SL(2,\mathbb{R})$ Wess-Zumino-Witten (WZW) model that is suitable to compute scattering amplitudes of winding strings in 3-dimensional Anti-de Sitter space at tree-level and one-loop. This is a refined version of previously proposed free-field realizations that, among other features, accomplishes to make the $H_3^+$ WZW-Liouville correspondence manifest.   

\end{abstract}

\newpage

\section{Introduction}

Understanding the physics of winding string states in three-dimensional Anti-de Sitter (AdS$_3$) Neveu-Schwarz--Neveu-Schwarz (NS-NS) backgrounds was crucial to establish the consistency of the theory at quantum level. The unitarity of the spectrum \cite{MO1}, the modular invariance of the one-loop partition function \cite{MO2}, and the consistency of scattering amplitudes \cite{MO3} were not well understood until the spectrally flowed sectors of the Hilbert space, associated to winding degrees of freedom, were discovered by Maldacena and Ooguri in Ref. \cite{MO1}. In the context of AdS/CFT correspondence, winding sectors were also very important to complete the dictionary between bulk and boundary three-point correlation functions \cite{Rastelo, CN}. 

The string scattering amplitudes involving winding string states in AdS$_3$ were first computed in Ref. \cite{GN3} by means of the Coulomb gas representation proposed in Ref. \cite{GN2}. The result was later reproduced by other methods \cite{MO3, succint, R}. One of the salient features of these scattering processes has to be with the conservation of the total winding number: Being not of topological origin, but due to the presence of the $B$-field in the background, the string winding number in AdS$_3$ space is not necessarily conserved by interactions. Nevertheless, this non-conservation exhibits a peculiar feature: As originally observed by Fateev, Zamolodchikov, and Zamolodchikov, the violation of the winding number in a tree-level amplitude is bounded from above. More precisely, for an $n$-string scattering amplitude to be non-vanishing, the total winding number $\Delta \omega = \omega_1 + \omega _2 + ... \omega_n$ has to obey $|\Delta \omega |\leq n-2$. The purpose of this paper is to present a formalism to compute tree-level and one-loop $n$-string winding violating scattering amplitudes in AdS$_3$ NS-NS background. This formalism is a refined version of the one proposed in \cite{GN2, GN3}, extending the latter in two directions: On the one hand, the prescription we propose here is suitable for describing one-loop amplitudes; namely, to compute Wess-Zumino-Witten (WZW) correlation functions on the torus. On the other hand, this version of the Coulomb gas realization accomplishes to make the correspondence between WZW and Liouville correlation functions manifest, and it represents an advantage.    

The paper is organized as follows: In Section 2, after a brief introduction to string theory in AdS$_3$ NS-NS backgrounds, we will study the worldsheet conformal field theory, which corresponds to the $H^+_3=SL(2,\mathbb{C})/SU(2)$ WZW theory. We will describe in detail the free field approach of the theory as originally proposed in \cite{GN2, GN3}. In Section 3, we will review the so-called $H_3^+$ WZW-Liouville correspondence. This is a remarkable tool that permits to write the string amplitudes in AdS$_3$ as a convolution of correlation functions of Liouville field theory. This will serve us as guideline in the construction. In Section 4, we will discuss the tree-level string scattering amplitudes. This will allow us to present our Coulomb gas formalism in a constructive way, first considering the maximally winding violating amplitudes, then analyzing the next-to-maximally winding violating process, and so on. In Section 5, we will discuss the one-loop scattering amplitudes. We will adapt our Coulomb gas method to the case of genus-1 Riemann surfaces, discussing the characteristic features of it, such as the existence of twisted sectors.

\section{String theory in AdS$_{3}$}

The non-linear $\sigma $-model describing the dynamics of strings in Euclidean AdS$_{3}$ space in presence of a NS-NS $B$-field corresponds to the WZW theory formulated on $H^+_3$. This permits to have access to the worldsheet formulation and solve the theory at quantum level. 

If Poincar\'e coordinates are used to describe AdS$_3$ space, the metric in the accessible patch takes the form 
\begin{equation}
ds^2 = \ell^2 (d\phi^2 + e^{2\phi }d\gamma d\bar{\gamma }), \label{Gmunu}
\end{equation}
where $\ell $ is the radius of the spacetime. The near boundary region of AdS$_3$ corresponds to large values of $\phi $. In these coordinates, the $B$-field configuration needed to support this background takes the form
\begin{equation}
B= \ell^2  e^{2\phi }d\gamma \wedge d\bar{\gamma }. \label{Bmunu}
\end{equation}

The relevant parameter in the story is the ratio between the radius of the spacetime and the string length scale; namely 
\begin{equation}
k=\frac{\ell^2}{\alpha '}  ,
\end{equation}
which controls the classical limit of the worldsheet theory, large $k$ corresponding to the semiclassical limit $\ell^2 > \ell^2_s = \alpha '$. The value of $k$ corresponds to the level of the WZW action.

Introducing auxiliary fields $\beta $, $\bar{\beta}$ and taking into account quantum corrections, the worldsheet string action on the background (\ref{Gmunu})-(\ref{Bmunu}) takes the form \cite{GKS}
\begin{equation}
S_{M}=\frac{1}{2\pi }\int d^{2}z\left( \partial \phi \bar{\partial }%
\phi +\beta \bar{\partial }\gamma +\bar{\beta }\partial \bar{%
\gamma }-\frac{R\phi }{2\sqrt{2(k-2)}}-2\pi M\ \beta \bar{\beta }e^{-%
\sqrt{\frac{2}{k-2}}\phi }\right) .  \label{sigma}
\end{equation}%

In these coordinates, the worldsheet CFT action involves a scalar field $\phi \in \mathbb{R}_{\geq 0}$ and a $\beta $-$\gamma $ commuting
ghost system. The term linear in $\phi $ couples to the two-dimensional curvature $R$ of the worldsheet; it introduces a background charge and, in the string theory language, it represents a linear dilaton configuration
\begin{equation}
\Phi (\phi) = \frac{\phi }{\sqrt{2(k-2)}} . \label{dilatonto}
\end{equation}

The factor $(k-2)^{-1/2}$ in the dilaton term comes from quantum corrections. In the path integral approach, this can be seen to arise when integrating over $\beta$, $\bar{\beta}$. This produces a kinetic piece $-\frac{2}{\pi}\int d^2z \partial\phi \bar{\partial} \phi $ and a background charge term $\frac{1}{4\pi }\int d^2zR\phi $ in the effective action \cite{Satoh1, Satoh2}. This results in (\ref{sigma}) by simply rescaling $\phi $. The integration over $\beta $, $\bar{\beta }$ also produces a factor $|\det \partial\bar{\partial}|^{-1}$. On the Riemann sphere, this factor can be absorbed in the normalization of the partition function. On the torus, because of the existence of twisting sectors and the dependence on the modular parameter, this factor deserves special attention \cite{Gawedzky}.

The value of the constant $M$ in (\ref{sigma}) can be set to 1 by shifting $\phi \to \phi + \sqrt{\frac{k-2}{2}} \log M $. This is, in fact, shifting the zero-mode of the dilaton (\ref{dilatonto}). Therefore, one may associate $M$ with the string coupling constant, namely 
\begin{equation}
M=g_s^2.
\end{equation}

Since the theory (\ref{sigma}) corresponds to the WZW action, it is known to exhibit $sl(2)_{k} \oplus sl(2)_{k}$ affine Kac-Moody symmetry. This algebra admits a free field realization, given by the so-called Wakimoto
representation. This amounts to define the local currents \cite{Wakimoto}
\begin{eqnarray}
J^{+}(z) &=&\beta (z),  \label{J1} \\
J^{3}(z) &=&-\beta (z)\gamma (z)-\sqrt{\frac{k-2}{2}}\partial \phi (z),
\label{J2} \\
J^{-}(z) &=&\beta (z)\gamma ^{2}(z)+\sqrt{2k-4}\gamma (z)\partial \phi
(z)+k\partial \gamma (z),  \label{J3}
\end{eqnarray}%
together with their complex conjugate counterparts $\bar{J}^3(\bar{z})$, $\bar{J}^{\pm }(\bar{z})$. Considering the free-field
propagators both for the scalar field $\phi $ and for the $\beta $-$\gamma $
system, namely
\begin{equation}
\left\langle \phi (z)\phi (w)\right\rangle =-\log (z-w)\text{%
,\quad }\left\langle \beta (z)\gamma (w)\right\rangle =\frac{1}{(z-w)} ,  \label{propagators}
\end{equation}
the operator product expansion (OPE) of currents (\ref{J1})-(\ref%
{J3}) realizes the $sl(2)_{k}$ affine Kac-Moody algebra in the following way
\begin{eqnarray}
J^{3}(z)J^{\pm }(w) &\simeq &\pm \frac{J^{\pm }(w)}{(z-w)}+\mathcal{O}(1) \label{U37} \\
J^{+}(z)J^{-}(w) &\simeq &\frac{k}{(z-w)^{2}}+\frac{2\ J^{3}(w)}{(z-w)}+%
\mathcal{O}(1) \\
J^{3}(z)J^{3}(w) &\simeq &\frac{-k/2}{(z-w)^{2}}+\mathcal{O}(1) \label{U39}
\end{eqnarray}%
where $\mathcal{O}(1)$ stands for regular terms. The interaction term in (\ref{sigma}) has a regular OPE with the currents, so it preserves the full $sl(2)_{k} \oplus sl(2)_{k}$ symmetry.

Through the Sugawara construction, the stress-tensor of the theory can be obtained from the currents (\ref{J1})-(\ref{J3}). This yields
\begin{equation}
T_{SL(2,\mathbb{R})}=\beta (z)\partial \gamma (z)-\frac{1}{2}\partial \phi
(z)\partial \phi
(z)-\frac{1}{\sqrt{2(k-2)}}\partial ^{2}\phi (z),  \label{pseudoT}
\end{equation}%
together with its anti-holomorphic counterpart $\bar{T}(\bar{z})$. The central charge of the worldsheet theory is
\begin{equation}
c_{WZW}=\frac{3k}{k-3} ,\label{cwert}
\end{equation}
which, as expected, tends to $3$ in the large $k$ limit. When the theory is formulated on AdS$_3 \times {\mathcal N}$ space, the value (\ref{cwert}) is complemented with the central charge of the internal CFT that describes the worldsheet theory on ${\mathcal N}$.

The spectrum of the theory is defined by Virasoro highest-height representations. These representations are build from Kac-Moody Verma moduli. Hermitian integrable representations for the $H^+_3$ WZW model are labeled by isospin variables $\left\vert j,m \right>$ with
\begin{equation}
j=-\frac{1}{2}+is , \ \ s\in \mathbb{R}_{>0}; \ \ \ m\in \mathbb{R} . \label{continua}
\end{equation} 

Starting from these states one constructs the spectrum of the Euclidean theory. The string theory in Lorentzian AdS$_3$ corresponds, instead, to the WZW model on $SL(2,\mathbb{R})$. The latter can be defined by analytic continuation from the $H^+_3$ model. This amounts to consider, in addition to the continuous series (\ref{continua}), the $SL(2,\mathbb{R})$ discrete representations\footnote{To be precise, since in order to avoid closed timelike curves here we are considering the universal covering of AdS$_3$, the representations to be considered are those of the universal covering of $SL(2,\mathbb{R})$, and in particular this implies that $j$ is not restricted to take semi-integer values.} 
\begin{equation}
j\in \mathbb{R}; \ \ \ m=-j+n, \ \ n\in \mathbb{Z}_{\geq 0} . \label{discreta1}
\end{equation}
and
\begin{equation}
j\in \mathbb{R}; \ \ \ m=j-n, \ \ n\in \mathbb{Z}_{\geq 0} . \label{discreta2}
\end{equation}

In terms of $j$, $m$, and $\bar{m}$ the angular momentum and the kinetic energy of strings in AdS$_3$ are given by $m-\bar{m}$ and $m+\bar{m}$, respectively. However, these are not all the relevant physical quantities that is necessary to consider to describe the string spectrum. A crucial step to define the theory was the observation of the necessity of including infinite new representations, which are constructed from the ones described above by acting with spectral flow automorphism of $sl(2)_k$ affine algebra \cite{MO1}. These new representations are labeled by an additional integer parameter $\omega $, whose physical interpretation is that of the winding number of the strings\footnote{Strictly speaking, this interpretation is accurate for the so-called ``long strings'', which correspond to the states of the continuous representations. For such states, in contrast with the ``short strings'' (\ref{discreta1})-(\ref{discreta2}), one has a notion of asymptotic states and, consequently, of S-matrix.}.

As said before, this winding number is not topological, but it is due to the presence of the NS-NS $B$-field (\ref{Bmunu}). This implies that the total winding number does not have to be necessarily conserved in a generic scattering process. A remarkable result is that, in a tree-level process involving $n$ strings in AdS$_3$, the total winding number $\Delta \omega = \omega_1 + \omega_2 + ... +\omega_n $ can indeed be violated, but its violation is bounded by $|\Delta \omega |\leq n-2$. This result had been originally observed for the $SL(2,\mathbb{R})/U(1)$ WZW model in an unpublished paper \cite{FZZ}, and it was later explained in Ref. \cite{MO3} in terms of the symmetries of the theory. The scattering process violating winding number have been first computed in Ref. \cite{GN3}, and the result was later re-derived by different methods in Refs. \cite{FH, MO3, succint, R}. In Ref. \cite{R}, it has been argued that $n$-string tree-level scattering processes in AdS$_3$ that violate the winding number in $|\Delta \omega |$ units can be expressed in terms of a convolution of $(2n-2+|\Delta \omega |)$-point functions of Liouville field theory on the sphere. The case $|\Delta \omega | <n-2$ was proven in \cite{R} and an educated conjecture has been made for the case $|\Delta \omega |=n-2$. The latter case was later proven in Ref. \cite{Gaston} by means of the Coulomb gas formalism proposed in \cite{GN2}. This formalism, which is similar to the one we will consider in this paper, consists of describing the theory on AdS$_3$ in terms of the WZW model on the product $SL(2,\mathbb{R})/U(1) \times U(1)$. While the $SL(2,\mathbb{R})/U(1) $ coset theory describes string theory on the Euclidean black hole background \cite{Witten}, the extra timelike direction $U(1)$ describes the time of the Lorentzian geometry. The idea of describing spectrally flowed sectors of the theory by modding-out a $U(1)$ current and subsequently adding it back, mimics the realization of spectral flow in ${\mathcal N}=2$ super-conformal algebra, and in the context of AdS$_3$ string theory was first suggested in \cite{MO1}. Here, we will consider the Coulomb gas formalism proposed in \cite{GN2} and further extend it. Let us begin by describing the theory on the coset in the next subsection.

\subsection{The coset construction}

The coset $SL(2,\mathbb{R)}/U(1)$ construction can be accomplished by
supplementing the $SL(2,\mathbb{R})$ model by adding an extra scalar field $%
X(z)=X_{L}(z)+X_{R}(\bar{z})$ and a $c=-2$ fermionic $B$-$C$ ghost system \cite{BK,DVV,Becker}.
This amounts to supplement the stress-tensor (\ref{pseudoT}) with two extra
pieces, namely 
\begin{equation}
T_{\text{coset}}=T_{SL(2,\mathbb{R})}-B(z)\partial C(z)-\frac{1}{2}%
\partial X(z)\partial X(z),  \label{T}
\end{equation}%
with 
\begin{equation}
\left\langle
X(z)X(w)\right\rangle =- \log(z-w) ,
\end{equation}
and analogously for the anti-holomorphic counterpart. This yields the central charge
\begin{equation*}
c_{\text{coset}}=\frac{3k}{k-2}-1.
\end{equation*}%

The BRST charge associated to the $U(1)$ of the coset is%
\begin{equation}
Q_{\text{BRST}}^{U(1)}=\int dz\ C(z)\left( J^{3}(z)-i\sqrt{\frac{k}{2}}\partial
X(z)\right) .  \label{BRST}
\end{equation}%

This means that the vertex operators creating physical states of the theory
on the coset have to have regular OPE with the current $J^{3}-i\sqrt{k/2}\partial X$.
Such operators are%
\begin{equation}
V^{\text{coset}}_{j,m,\bar{m}}(z,\bar{z})=\gamma ^{j-m}(z)\bar{\gamma }^{j-%
\bar{m}}(\bar{z})e^{\sqrt{\frac{2}{k-2}}j\phi (z,\bar{z})}\times e^{i%
\sqrt{\frac{2}{k}}(mX_{L}(z)+\bar{m}X_{R}(\bar{z}))},  \label{V}
\end{equation}%
where $j$, $m$ and $\bar{m}$ are isospin variables that label the $SL(2,%
\mathbb{R})\times SL(2,\mathbb{R})$ representations. The conformal dimension of these operators is
\begin{equation}
h_{j,m}^{\text{coset}} = -\frac{j(j+1)}{k-2}+\frac{m^2}{k}, \ \ \ \ \ \bar{h}_{j,\bar{m}}^{\text{coset}} = -\frac{j(j+1)}{k-2}+\frac{\bar{m}^2}{k} .
\end{equation}

\subsection{Adding the spectrally flowed sectors}

Spectral flow automorphism of $sl(2)_k$ affine algebra generates new representations \cite{MO1}. The Virasoro primary states corresponding to the spectrally flowed sectors of the spectrum represent the winding strings, characterized by the extra quantum number $\omega \in \mathbb{Z}$. A free field realization of such states was achieved by considering the theory on the product $SL(2,\mathbb{R})/U(1) \times U(1)$. This amounts to add to the coset construction discussed above an extra timelike free boson $T(z)$. Explicitly, one improves the currents as\footnote{Hereafter, $T(z)$ will represent a timelike free boson. It is important not to mistake it for the holomorphic component of the stress-tensor.}
\begin{eqnarray}
J^{\pm }(z) &\to & J^{\pm }(z) \ e^{\pm i \sqrt{\frac{2}{k}} (X(z)+T(z))} \\
J^{3} (z) &\to & J^3(z)-i\sqrt{\frac{k}{2}} (\partial X(z) +\partial T(z) ) \ , \label{323}
\end{eqnarray}
with%
\begin{equation}
\left\langle T(z)T(w)\right\rangle =+\log (z-w)
\end{equation}%
and its anti-holomorphic counterpart. These currents reproduce the OPE (\ref{U37})-(\ref{U39}).

Now, let us compose the theory on the product $SL(2,\mathbb{R})/U(1)\times U(1)$ and study the spectrum. The states of the theory are given by representations of $SL(2,\mathbb{R})\times SL(2,\mathbb{R})$, including the spectral flow quantum number $\omega $. Let us denote these states by $\left\vert j,m,\bar{m},\omega \right\rangle $. They are created by operators
\begin{equation}
V _{j,m,\bar{m}}^{\omega }(z)=\gamma ^{j-m}(z)e^{\sqrt{\frac{2%
}{k-2}}j\phi (z)}e^{i\sqrt{\frac{2}{k}}mX(z)}e^{i\sqrt{\frac{2}{k}}(m+\frac{k%
}{2}\omega )T(z)}\times h.c.  \label{Vertex}
\end{equation}%
where $h.c.$ stands for the anti-holomorphic part\footnote{Strictly speaking, the part that is omitted depends on the variable $\bar{z}$ and the momentum $\bar{m}$.}. These operators are the product of operators (\ref{V}) and the exponential operators that carry the charge under $T(z)$. This charge corresponds to the total energy $E=m+\bar{m}+k\omega $. Unlike $\omega $, the energy is conserved in the scattering processes.

Operators (\ref{Vertex}) create the \textit{in}-states $\left\vert j,m,%
\bar{m},\omega \right\rangle $ from the $SL(2,\mathbb{R})$ invariant
vacuum $\left\vert 0\right\rangle $; namely%
\begin{equation}
\lim_{z\rightarrow 0}V _{j,m,\bar{m}}^{\omega }(z,\bar{z})\left\vert
0\right\rangle =\left\vert j,m,\bar{m},\omega \right\rangle .
\label{in}
\end{equation}

The \textit{out}-states are given by%
\begin{equation}
\left\langle j,m,\bar{m},\omega \right\vert =\lim_{z\rightarrow \infty }%
\text{ }z^{2h_{j,m}^{\omega }}\bar{z}^{2\bar{h}_{j,\bar{m}}^{\omega }}%
\text{ }\left\langle 0\right\vert V _{-1-j,m,\bar{m}}^{\omega }(z,\bar{z}),
\label{out}
\end{equation}%
where $h_{j,m}^{\omega }$ is the value of the conformal dimension of $V
_{j,m,\bar{m}}^{\omega }$, given by
\begin{equation}
h^{\omega }_{j,m} = h^{\text{coset} }_{j,m} -\frac{(m+k\omega/2)^2}{k} = -\frac{j(j+1)}{k-2} - m\omega -\frac{k}{4}\omega ^2 \ , \ \ \ \ \bar{h}^{\omega }_{j,\bar{m}} = {h}^{\omega }_{j,\bar{m}} \label{h} .
\end{equation}

The OPE of the vertex operators (\ref{Vertex}) and the currents is
\begin{eqnarray}
J^{3}(z)V _{j,m,\bar{m}}^{\omega }(w) &\simeq &\frac{(m+k\omega /2)}{%
(z-w)}V _{j,m,\bar{m}}^{\omega }(w)+\mathcal{O}(1)  \label{j3} \label{mpj} \\
J^{\pm }(z)V _{j,m,\bar{m}}^{\omega }(w) &\simeq &\frac{(\pm j-m)}{%
(z-w)^{1\pm \omega }}V _{j,m\pm 1,\bar{m}}^{\omega }(w)+\mathcal{O}%
((z-w)^{\mp \omega }) . \label{jpm}
\end{eqnarray}

This means that states $\left\vert j,m,\bar{m},\omega \right\rangle $ are also eigenstates of the $J^3(z)$, $\bar{J}^3(\bar{z})$ currents (\ref{323}), with eigenvalues
\begin{equation}
p_L = m+\frac{k}{2}\omega \ , \ \ \ \ \ p_R = \bar{m}+\frac{k}{2}\omega , \label{329bis}
\end{equation}
respectively. It is worthwhile noticing that formulas (\ref{h}) and (\ref{329bis}) are both symmetric under the Weyl reflection
\begin{equation}
(j,m,\bar{m},\omega ) \ \ \to \ \ (-1-j,m,\bar{m},\omega ) 
\end{equation}
and the reflection
\begin{equation}
(j,m,\bar{m},\omega ) \ \ \to \ \ (j,-m,-\bar{m},-\omega ) .
\end{equation}

These $\mathbb{Z}_2$ symmetries allow us to consider here correlation functions with $\omega_1 +\omega_2 + ... + \omega_n \leq 0$ without loss of generality.

\subsection{Conjugate vertex representations}

Apart from the Wakimoto type representation (\ref{Vertex}), to which we will refer as ``standard representation'', there exists another free-field representation of Virasoro primaries with conformal dimension (\ref{h}) and OPE (\ref{mpj})-(\ref{jpm}). This second representation is constructed by considering in the coset $SL(2,\mathbb{R})/U(1)$ the free field realization of discrete states proposed in Ref. \cite{BK} and adding to it the charge under the field $T(z)$. Explicitly, the operators take the form
\begin{equation}
\tilde{V }_{j,m,\bar{m}}^{\omega }(z)=\frac{1}{Z_{j,m}}\beta
^{j+m}(z)e^{-\sqrt{\frac{2}{k-2}}(j+\frac{k}{2})\phi (z)}e^{i\sqrt{\frac{2}{k%
}}(m-\frac{k}{2})X(z)}e^{i\sqrt{\frac{2}{k}}(m+\frac{k}{2}\omega
)T(z)}\times h.c.  \label{conjugated}
\end{equation}%
where $(Z_{j,m}\bar{Z}_{j,\bar{m}})^{-1}$ is a normalization factor to be conveniently fixed. 

The $\beta $-dependent operators (\ref{conjugated}) are reminiscent of the Dotsenko conjugated representation of the $SU(2)$ WZW theory \cite{Dotsenko1, Dotsenko2}. They create conjugate \textit{in}-states from a conjugate vacuum $\left\vert \tilde{0}\right> $; namely,%
\begin{equation}
\lim_{z\rightarrow 0}\tilde{V }_{j,m,\bar{m}}^{\omega
}(z,\bar{z})\left\vert \tilde{0}\right> =\left\vert j_{n},m_{n},\bar{m}_{n},\omega
_{n}\right\rangle .  \label{inconjugated}
\end{equation}

An standard way of fixing the normalization $Z_{j,m}^{-1}$ in (\ref{conjugated}) is requiring the 2-point function to be normalized as
\begin{equation}
\left\langle j,m,\bar{m},\omega \right. \left\vert j,-m,-\bar{m}%
,-\omega \right\rangle \equiv 1. \label{mmmmmmJ}
\end{equation}%

It is possible to verify that, with this normalization, the OPE with
the currents also gives
\begin{eqnarray}
J^{3}(z)\tilde{V }_{-1-j,m,\bar{m}}^{\omega }(w) &\simeq &\frac{%
(m+k\omega /2)}{(z-w)}\tilde{V }_{j,m,\bar{m}}^{\omega }(w)+%
\mathcal{O}(1) \\
J^{\pm }(z)\tilde{V }_{-1-j,m,\bar{m}}^{\omega }(w) &\simeq &%
\frac{(\pm j-m)}{(z-w)^{1\pm \omega }}\tilde{V }_{-1-j,m\pm 1,%
\bar{m}}^{\omega }(w)+\mathcal{O}((z-w)^{\mp \omega }),
\end{eqnarray}%
which coincides with (\ref{j3})-(\ref{jpm}). That is, operators $V _{-1-j,m,\bar{m}}^{\omega }$ and 
$\tilde{V }_{j,m,\bar{m}}^{\omega }$ create states with the same
conformal dimension and the same OPE with the currents. However, normalization (\ref{mmmmmmJ}) is not the only solution that yields the OPE (\ref{jpm})-(\ref{mpj}). In fact, one can also consider
\begin{equation}
{Z_{j,m}} {\bar{Z}_{j,\bar{m}}} = (-1)^{j+m} c\frac{\Gamma (1+j+m)}{\Gamma (-j-\bar{m})} , \label{uuuuuu}
\end{equation}
with $c$ being a constant independent of $j$, $m$, and $\bar{m}$. We will adopt (\ref{uuuuuu}) for convenience.

For computing correlation functions in the Coulomb gas (free field) approach, it is crucial the inclusion of screening operators \cite{Becker, GN2, GN3}. These are operators of conformal dimension $(1,1)$ that, in addition, have regular OPE with the $sl(2)_k$ currents (up to a total derivative).

The screening operators of the $SL(2,\mathbb{R})$ WZW theory are
\begin{eqnarray}
S_{+} (z) = c \ \tilde{V}_{1-\frac{k}{2},\frac{k}{2},\frac{k}{2}}^{-1} = \beta (z)  e^{-\sqrt{\frac{2}{k-2}} \phi(z)} \times h.c. , \label{Smas}
\end{eqnarray}
and
\begin{eqnarray}
S_{-} (z) = c \ \tilde{V}_{\frac{k}{2}-2,\frac{k}{2},\frac{k}{2}}^{-1} = \beta^{k-2} (z)  e^{-\sqrt{2({k-2})} \phi(z)} \times h.c. \label{Smenos}
\end{eqnarray}

While (\ref{Smas}) corresponds to the interaction term appearing in (\ref{sigma}) and, therefore, can be understood simply in terms of the path integral approach to the $\sigma $-model, the physical interpretation of dual screening operator (\ref{Smenos}) is less clear. The latter can be associated to worldsheet instanton effects. In Ref. \cite{GN3}, it has been shown that operator (\ref{Smenos}) can be consistently used to compute WZW correlation functions, leading to the exact result. Here, instead, we will consider only the usual screening operator (\ref{Smas}).

The theory also contains other special operators. Such is the case of the dimension-$(0,0)$ operator
\begin{equation}
\tilde{V }_{0,0,0}^{0}(z)=e^{-\sqrt{\frac{2}{k-2}}\frac{k}{2}\phi
(z)}e^{-i\sqrt{\frac{k}{2}}X(z)}\times h.c.  \label{identity}
\end{equation}
which can also be associated to representations (\ref{Vertex}) as follows
\begin{equation}
V _{-\frac{k}{2},-\frac{k}{2},-\frac{k}{2}}^{1}(z)=e^{-\sqrt{\frac{2}{k-2}%
}\frac{k}{2}\phi (z)}e^{-i\sqrt{\frac{k}{2}}X(z)}\times h.c.
\end{equation}

This operator has been dubbed in Ref. \cite{MO3} ``spectral
flow operator'', and, following the prescription originally proposed in
Ref. \cite{FZZ}, it was used to compute correlation functions that violate winding
number conservation. In the unpublished paper \cite{FZZ}, operator (\ref{identity}) was referred to as ``conjugated representation of the identity operator''.

Other interesting objects are the dimension-$(1,1)$ operators 
\begin{eqnarray}
 \tilde{V}_{-1,0,0}^{0} (z) &=&  c^{-1/2} \beta^{-1}(z) e^{-\sqrt{\frac{k-2}{2}}\phi (z) } e^{-i\sqrt{\frac{k}{2}}X(z)} \times h.c., \\
 {V}_{1-\frac{k}{2},\frac{k}{2},\frac{k}{2}}^{-1 } (z) &=&  \gamma(z) 
 e^{-\sqrt{\frac{k-2}{2}}\phi(z)} e^{i\sqrt{\frac{k}{2}} X(z)} \times h.c.
\end{eqnarray}
and
\begin{equation}
\tilde{V }_{1-k,k,k}^{-2}(x) = c^{-1/2} \beta (x)e^{\sqrt{\frac{k-2}{2}}\phi (x)}e^{i\sqrt{\frac{k}{2}%
}X(x)}\times h.c.  \label{elpibe}
\end{equation}%

The latter will be very important in our discussion later. 

\subsection{Correlation functions}

We are interested in computing $n$-point correlation functions of Virasoro primary operators in the WZW theory. Formaly, these observables are the expectation values
\begin{equation}
X_{n}^{\Delta \omega }\equiv \left\langle \prod_{i=1}^{n} V_{j_i,m_i,\bar{m}_i}^{\omega_i } (z_i) \right\rangle_{\text{WZW}} , \label{cf}
\end{equation}
where here the superscript $\Delta \omega $ refers to the total winding number in the correlators\footnote{We omit the dependence on other variables such as $j_i$, $m_i$, $\bar{m}_i$, $z_i$, $\bar{z}_i$.}.

We will follow the prescription for computing correlation functions presented in Ref. \cite{GN2} and eventually propose a refined version of it that is suitable for the genus-1 generalization. The prescription of \cite{GN2} is the following: Consider the correlation function
\begin{equation}
X_{n}^{\Delta \omega }=\ \left\langle j_{1},m_{1},\bar{m}_{1},\omega
_{1}\right\vert \prod_{t=2}^{p}V _{-1-j_{t},m_{t},\bar{m}%
_{t}}^{\omega _{t}}(z_{t})\prod_{l=p+1}^{n-1}\tilde{V }%
_{j_{l},m_{l},\bar{m}_{l}}^{\omega _{l}}(z_{l})\left\vert j_{n},m_{n},%
\bar{m}_{n},\omega _{n}\right> ,   \label{LA249}
\end{equation}%
which involves both fields of the standard representation (\ref{Vertex}) and the conjugate representation (\ref{conjugated}). The first question is which representation has to be used to describe the {\it in}-state $\left\vert j,m,\bar{m},\omega \right>$, i.e. whether one has to resort to the standard representation (\ref{Vertex}) or to the conjugated representation (\ref{conjugated}). The bottom line of the next subsection will be that, in fact, one can actually make use of any of these representations. The only detail to be taken into account is that, depending on which representation one decides to use, the charge compensation condition to consider has to be consistent with such choice.

We define the {\it out}-state in the correlator as
\begin{equation}
\left\langle j_1,m_1,\bar{m}_1,\omega_1
\right\vert = \lim _{z_1 \to \infty } z_1^{2h^{\omega_1 }_{j_1,m_1}} \bar{z}_1^{2\bar{h}^{\omega_1 }_{j_1,\bar{m}_1}} \left\langle 0
\right\vert V_{-1-j_1,m_1,\bar{m}_1}^{\omega_1 }(z_1 , \bar{z}_1) 
\end{equation}
and the {\it in}-state as
\begin{equation}
\left\vert j_n,m_n,\bar{m}_n,\omega_n \right> = \lim_{z_n\to 0} \tilde{V}_{j_n,m_n,\bar{m}_n}^{\omega_n }(z_n, \bar{z}_n ) \left\vert \tilde{0} \right> , \label{jhgf}
\end{equation}
or alternatively as in (\ref{in}); see subsection 2.5.

In the Coulomb gas approach \cite{Becker, GN3}, the correlators involve, in addition, $s$ integrated screening operators $S_{+}$; more precisely, the correlators take the form
\begin{eqnarray}
X_{n}^{\Delta \omega }= \Gamma(-s) c^{s} g_s^{2s} \int \prod_{r=1}^{s}d^2w_r \left\langle    \prod_{t=1}^{p}V _{-1-j_{t},m_{t},\bar{m}%
_{t}}^{\omega _{t}}(z_{t})\prod_{l=p+1}^{n}\tilde{V }%
_{j_{l},m_{l},\bar{m}_{l}}^{\omega _{l}}(z_{l})  \prod_{r=1}^{s} \tilde{V}_{1-\frac{k}{2},\frac{k}{2},\frac{k}{2}}^{-1}(w_r)
\right>_{M=0}  \\ \label{setitfri}
\end{eqnarray}%
where $s=-\sum_{i=1}^{n}j_i - p-\frac{k}{2}(n-p+1)$, and where the subscript $M=0$ refers to the fact that the correlator (\ref{setitfri}) is defined in terms of the free action, i.e. $M =0$ in (\ref{sigma}).

Charge compensation conditions in (\ref{setitfri}) are defined by demanding the fields in the correlators to equal the charges of the conjugate identity operator (\ref{identity}), \cite{Dotsenko1, Dotsenko2}. In particular, if one considers (\ref{jhgf}), the compensation of the charge associated to the field $\phi (z)$ demands
\begin{equation}
\frac{k}{2}(p-n)-p- s-\sum_{i=1}^{n}j_i =-\frac{k}{2} , \label{AlA}
\end{equation}
which follows from the fact that $\tilde{V}_{0,0,0}^{0}(z)$ has charge $-k/2$ under $\phi (z)$. Analogously, the compensation of the charge associated to the field $X (z)$ reads
\begin{equation}
\sum_{i=1}^{n}m_i +\frac{k}{2}(p-n)=\sum_{i=1}^{n}\bar{m}_i +\frac{k}{2}(p-n)=-\frac{k}{2}, \label{AAlA}
\end{equation}
expressing the fact that the charge of $\tilde{V}^0_{0,0,0}$ associated to that field is also $-k/2$. The compensation of the charge associated to the field $T (z)$, which expresses the energy conservation, reads
\begin{equation}
\sum_{i=1}^{n}(m_i +\frac{k}{2}\omega_i)=\sum_{i=1}^{n}(\bar{m}_i +\frac{k}{2}\omega_i) = 0 , \label{352}
\end{equation}
and, then, equations (\ref{AlA})-(\ref{352}) imply the total winging number violation
\begin{equation}
\sum_{i=1}^n \omega_i = p+1-n . \label{AlZ}
\end{equation}

That is, the total winding number (non-)conservation in (\ref{setitfri}) depends on $p$, i.e. on the relative amount of operators in the representation (\ref{Vertex}) or in the representation (\ref{conjugated}) one decides to include in the correlator. Such is the prescription proposed in Ref. \cite{GN2}, which has been used in Ref. \cite{GN3} to compute both winding preserving and winding violating string amplitudes. In Ref. \cite{Gaston}, the case $p=1$ with generic $n$, i.e. the maximally violating amplitude, has been studied in relation to a previous conjecture about the functional form of such observables \cite{R}. 

The charge compensation conditions (\ref{AlA})-(\ref{AlZ}) also yield a condition for the relative amount of $\beta $ and $\gamma$ ghost fields in (\ref{setitfri}). If one denotes by $N_{\beta } (X_{n}^{\Delta \omega })$ and $N_{\gamma } (X_{n}^{\Delta \omega })$ the amount of operators $\beta $ and $\gamma $ included in a given correlator $X_{n}^{\Delta \omega }$, respectively, this condition reads
\begin{equation}
N_{\beta} (X_{n}^{\Delta \omega }) - N_{\gamma }(X_{n}^{\Delta \omega }) = N_{\bar{\beta }}(X_{n}^{\Delta \omega }) - N_{\bar{\gamma }}(X_{n}^{\Delta \omega }) = 0 . \label{LA258}
\end{equation}
This follows from $N_{\beta }(\tilde{V}_{0,0,0}^0)=N_{\gamma }(\tilde{V}_{0,0,0}^0)=0$. Similar condition has been considered, for instance, in Ref. \cite{Andreev}. However, this is not the relation obtained from the path integral approach when integrating over the zero-modes of the ghost fields. In fact, one rather obtains \cite{Becker}
\begin{equation}
N_{\beta}(X_{n}^{\Delta \omega }) - N_{\gamma }(X_{n}^{\Delta \omega }) = N_{\bar{\beta }}(X_{n}^{\Delta \omega }) - N_{\bar{\gamma }}(X_{n}^{\Delta \omega }) = 1 . \label{355}
\end{equation}

Then, the question is how to find a prescription for computing amplitudes that, while being consistent with the method of \cite{GN2} and the result (\ref{AlZ}), is at the same time consistent with the condition (\ref{355}) arising in the path integral approach. This question is motivated by the fact that having such a path integral realization would be useful to extend the computation to higher loops. Such a prescription is actually possible and we will discuss it in the next subsection.

\subsection{Alternative representation and path integral} 

The main idea to obtain the condition (\ref{355}) within the prescription of Ref. \cite{GN2} is to describe the {\it in}-state $\left\vert j_n ,m_n ,\bar{m}_n ,\omega_n \right>$ in terms of the representation $V_{-1-j,m,\bar{m}}^{\omega }(z)$ instead of $\tilde{V }_{j,m,\bar{m}}^{\omega }(z)$; namely, consider
\begin{equation}
X_{n}^{\Delta \omega }=\lim_{z_{1}\rightarrow \infty }\text{ }%
z_{1}^{2h_{j_{1},m_{1}}^{\omega _{1}}}\bar{z}_{1}^{2\bar{h}_{j_{1},\bar{m%
}_{1}}^{\omega _{1}}}\left\langle  \prod_{t=1}^{p}V
_{-1-j_{t},m_{t},\bar{m}_{t}}^{\omega
_{t}}(z_{t})\prod_{l=p+1}^{n-1}\tilde{V }_{j_{l},m_{l},%
\bar{m}_{l}}^{\omega _{l}}(z_{l})V _{-1-j_{n},m_{n},\bar{m}%
_{n}}^{\omega _{n}}(0) \right\rangle_{\text{WZW}}   \label{Dos}
\end{equation}%
where, instead of defining the charge compensation condition with respect to the conjugate identity $\tilde{V }_{0,0,0}^0(z)$, we have to do it with respect to the Weyl-reflected version of the operator $V_{0,0,0}^0=1$; that is, with respect to
\begin{equation}
\hat{V }_{0,0,0}^{0}(z)\equiv V_{-1,0,0}^{0} = \gamma ^{-1}e^{-\sqrt{\frac{2}{k-2}}\phi }\times
h.c. \label{358}
\end{equation}

The first of these
charge compensation conditions is given by requiring the difference between
the amount of fields $\beta $ and fields $\gamma $ in the correlator $%
X_{n}^{\Delta \omega }$ to be equal to the difference between the amount of
fields $\beta $ and fields $\gamma $ in the conjugated identity operator (\ref{358}). That is,
\begin{equation}
N_{\beta }(X_{n}^{\Delta \omega })-N_{\gamma }(X_{n}^{\Delta \omega
})=N_{\beta }(\hat{V }_{0,0,0}^{0})-N_{\gamma }(\hat{V }%
_{0,0,0}^{0})=1  \label{A1}
\end{equation}%
since $N_{\beta }(\hat{V }_{0,0,0}^{0})=0$ and $N_{\gamma }(\hat{V }_{0,0,0}^{0})=-1$, cf. (\ref{LA258}). For correlators (\ref{Dos}) this condition
reads explicitly
\begin{equation}
\sum_{l=1}^{n}j_{l}+\sum_{l=1}^{n}m_{l}+p+s=0.  \label{B1}
\end{equation}%

The rest of the condition comes from equaling the charges
associated to fields $\phi (z)$, $X(z)$, and $T(z)$. Those are
\begin{equation}
\sum_{l=1}^{n}j_{l}+p+s+\frac{k}{2}(n-p-1)=0,  \label{C1}
\end{equation}
together with
\begin{equation}
\sum_{l=1}^{n}m_{l}=\frac{k}{2}(n-p-1),\qquad \sum_{l=1}^{n}m_{l}+\frac{k}{2}%
\sum_{l=1}^{n}\omega _{l}=0,  \label{D1}
\end{equation}
and
\begin{equation}
\sum_{l=1}^{n}\omega _{l}=p+1-n.  \label{E1}
\end{equation}

Notice that (\ref{B1})-(\ref{E1}) exactly agree with (\ref{AlA})-(\ref{AlZ}). Let us show that, in addition, conditions (\ref{B1})-(\ref{E1}) also agree with those obtained by integrating over the zero-modes of the fields in the path integral approach. In this approach, the correlation functions are defined as
\begin{equation}
X_n^{\Delta \omega} = \int {\mathcal D}^2\beta {\mathcal D}^2\gamma {\mathcal D}\phi \ e^{-S_M} 
\prod_{i=1}^{p+1} V_{-1-j_i, m_i, \bar{m}_i}^{\omega _i}(z_i)
\prod_{l=p+2}^{n} \tilde{V }_{j_l, m_l, \bar{m}_l}^{\omega _l}(z_l) .
\end{equation}

Since $\beta $, $\bar{\beta }$ are 1-differentials and the $\gamma $, $\bar{\gamma }$ are 0-differentials, the Riemann-Roch on the Riemann surface of genus-$g$ yields
\begin{equation}
N_{\beta }(X_{n}^{\Delta \omega })-N_{\gamma }(X_{n}^{\Delta \omega })=N_{\bar{\beta } }(X_{n}^{\Delta \omega })-N_{\bar{\gamma } }(X_{n}^{\Delta \omega })=1-g \ ,
\end{equation}
which reduces to (\ref{A1}) when $g=0$. On the other hand, the integration over the zero-mode $\phi_0$ of field $\phi (z)$, yields\footnote{Here, we have absorbed a $k$-dependent factor in the definition of the path integral. We also have made the replacement $z_n \leftrightarrow z_{p+1}$.}
\begin{eqnarray}
X_n^{\Delta \omega} &=& (-1)^s\Gamma(-s)c^2g_s^{2s}\int \prod_{r=1}^{s}  d^2w_r\int {\mathcal D}^2\beta {\mathcal D}^2\gamma {\mathcal D}\phi \ e^{-S_0} 
\prod_{i=1}^{p+1} V_{-1-j_i, m_i, \bar{m}_i}^{\omega _i}(z_i) \times \nonumber \\
&& \times \prod_{l=p+2}^{n} \tilde{V }_{j_l, m_l, \bar{m}_l}^{\omega _l}(z_l)  \prod_{r=1}^{s} \tilde{V }_{1-\frac{k}{2}, \frac{k}{2}, \frac{k}{2} }^{-1}(w_r) ,
\end{eqnarray}
where the operators are now evaluated on the fluctuations $\phi(z)-\phi_0$, the action $S_0$ corresponds to (\ref{sigma}) with $M=0$, and $s$ is given by
\begin{equation}
\sum_{l=1}^{n}j_{l}+p+s+1+\frac{k}{2}(n-p-1)=1-g.
\end{equation}%
This condition follows from the Gauss-Bonnet theorem on the sphere because of the existence of the background charge (\ref{dilatonto}), and in the case $g=0$ it coincides with (\ref{C1}). 

The integration over the zero-modes of fields $X(z)$ and $T(z)$ is substantially simpler and yields
\begin{equation}
\sum_{l=1}^{n}m_{l}=\frac{k}{2}(n-p-1),\qquad \sum_{l=1}^{n}m_{l}+\frac{k}{2}%
\sum_{l=1}^{n}\omega _{l}=0,
\end{equation}%
which implies%
\begin{equation}
\sum_{l=1}^{n}\omega _{l}=p+1-n,
\end{equation}
in exact agreement with (\ref{D1})-(\ref{E1}). Then, we have a path integral realization of correlation functions that yields $\Delta \omega \neq 0$.

\section{Liouville strings in AdS$_3$}

Besides the Coulomb gas prescription of \cite{GN2}, another important tool in our discussion will be a close relation that exists between correlation functions in the $H_3^+$ WZW theory and correlation functions in Liouville field theory. In the next subsection, we will briefly review the latter theory, and in subsection 3.2 we will review the precise correspondence between $H_3^+$ WZW and Liouville observables.

\subsection{Liouville field theory}

Liouville theory naturally arises in the formulation of the two-dimensional
quantum gravity and in the path integral quantization of string theory. It is also closely related to Einstein gravity in AdS$_3$ space, and to four-dimensional ${\mathcal N}=2$ superconformal gauge theories. This is a non-rational conformal field theory whose action reads%
\begin{equation}
S_{L}=\frac{1}{4\pi }\int d^{2}z\left( \partial \varphi \overline{%
\partial }\varphi +\frac{1}{2\sqrt{2}}QR\varphi +4\pi e^{\sqrt{2}%
b\varphi }\right) . \label{mancha}
\end{equation}%

The background charge parameter takes the
value $Q=b+b^{-1} $ in order to render the potential barrier $e^{%
\sqrt{2}b\varphi }$ an exact marginal operator. The central charge of Liouville theory is
given by%
\begin{equation}
c_{L}=1+6Q^{2}. \label{UFALACL}
\end{equation}%

Important objects of
the theory are the exponential vertex operators
\begin{equation}
V_{\alpha }(z)=e^{\sqrt{2}\alpha \varphi (z)}, \label{hgfd}
\end{equation}
which are local primary operators of conformal dimension 
\begin{equation}
h^{L}_{\alpha}=\bar{h}^{L}_{\alpha}=\alpha
(Q-\alpha ).
\end{equation}

A special case of operator (\ref{hgfd}) is the degenerate field of momentum $\alpha =-1/(2b)$,
\begin{equation}
V_{-\frac{1}{2b}}(x)=e^{-\frac{1}{\sqrt{2}b} \varphi (x)}, \label{degenerado}
\end{equation}
which creates a non-normalizable state of dimension $h_{-1/(2b)}<-1/2$ that is annihilated by the Virasoro operator $L_{-1}^2+b^{-2}L_{-2}$. That is, vertex operator (\ref{degenerado}) is a degenerate field in the sense that it contains a null state in the Verma modulo. 

Correlation functions in Liouville theory are formally defined as follows%
\begin{equation}
\left\langle \prod_{i=1}^{n} V_{\alpha
_{i}}(z_{i}) \right\rangle _{L}\equiv \int
{\mathcal D}\varphi \ e^{-S_{L}}\prod_{i=1}^{n}e^{\sqrt{2}\alpha _{i}\varphi
(z_{i})} . \label{verta}
\end{equation}

In the case that at least one of the $n$ operators in (\ref{verta}) has momentum $\alpha =-1/(2b)$, as in (\ref{degenerado}), the correlation function happens to solve the Belavin-Polyakov-Zamolodchikov (BPZ) equation. 

On the spherical topology, generic correlation functions (\ref{verta}) admit the integral representation \cite{GoulianLi}
\begin{equation}
\left\langle \prod_{i=1}^{n} V_{\alpha
_{i}}(z_{i}) \right\rangle_{\text{L}} = b^{-1}\Gamma (-s) \prod_{i<j}^n |z_i -z_j|^{-4\alpha_i\alpha_j} \int \prod_{r=1}^{s}d^2w_r \prod_{i=1}^{n}\prod_{r=1}^{s} |z_i-w_r|^{-4b\alpha_i} \prod_{r<t}^{s} |w_r-w_t|^{-4b^2} , \label{43}
\end{equation}
with $s$ given by
\begin{equation}
bs+\sum_{i=1}^{N}\alpha _{i}=Q.  \label{pupo2}
\end{equation}%

Expression (\ref{43}) makes sense for $s\in \mathbb{Z}_{\geq 0}$. However, it can be analytically continued to other (complex) values of $s$. For instance, in the case of the partition function, and the 2- and the 3-point functions ($n=0,2,3$), integral (\ref{43}) can be exactly solved in terms of the Dotsenko-Fateev formula \cite{DF1, DF2, DF3} and the result can be extended to complex values of $\alpha_i $ and $b$.

\subsection{The $H_3^+$ WZW - Liouville correspondence}

In Ref. \cite{RibaultTeschner}, a remarkable connection between the $H_3^+$ WZW correlation functions and Liouville correlation functions was derived. This is based on a previous work of Stoyanovsky \cite{Stoyanovsky}, who noticed a connection between the Knizhnik-Zamolodchikov and the BPZ equations. The correspondence between $H_3^+$ WZW and Liouville correlators, as presented in \cite{RibaultTeschner}, states that $n$-point functions in the WZW theory are given by a convolution $(2n-2)$-point function of Liouville field theory, the latter involving $n$ generic primary operators (\ref{verta}) and $n-2$ degenerate operators (\ref{degenerado}). The formula of \cite{RibaultTeschner} was re-derived in Ref. \cite{HikidaSchomerus} using the path integral approach, and then extended to arbitrary genus $g\geq 0$. In Ref. \cite{R}, Ribault proposed a generalization of the $g=0$ formula to the case in which spectrally flowed states are included in the WZW correlator. The result reads\footnote{Notice that, while the worldsheet central charge is given by (\ref{cwert}), the Liouville central charge (\ref{UFALACL}) in terms of $k$ reads $c_L=c_{WZW}-2+6k$. A curious feature is that in the semi-classical (large $k$) limit the Liouville central charge agrees with the central charge of the dual boundary CFT \cite{GKS, Ooguri}.}
\begin{eqnarray}
\left\langle \prod_{i=1}^{n} V_{j_i,m_i,\bar{m}_i}^{\omega_i } (z_i)\right>_{\text{WZW}} &=& \frac{2\pi^{3-2n}b c^{\Delta\omega }}{\Gamma 
(n-1+\Delta\omega)} \prod_{i=1}^{n} \frac{\Gamma(-j_i-m_i)}{\Gamma(j_i+1+\bar{m}_i)}  \prod_{i<j} |z_i - z_j|^{2\beta_{ik}} \nonumber  \times \\
&&\times \int\prod_{l=1}^{n-2-|\Delta\omega |} d^2x_l \prod_{i=1}^{n}\prod_{l=1}^{n-2-|\Delta\omega |} |z_i-x_l|^{2m_i-k} 
\prod_{l<t}^{n-2-|\Delta\omega |} 
|x_l - x_t|^{2k} \nonumber \times \\ 
&&\times \left\langle \prod_{i=1}^{n} V_{\alpha_i} (z_i) \prod_{l=1}^{n-2-|\Delta\omega |} V_{-\frac{1}{2b}} 
(x_l)\right>_{\text{L}} \label{Ribault}
\end{eqnarray}
where $\beta_{ik}=k-k\omega_i \omega_j -2\omega _i m_j -2 
m_i\omega_j-2m_i-2m_j$, $c$ is a constant independent of $j_i$, $m_i$, and $z_i$, and where\footnote{Here, we are considering the particular case $m_i=\bar{m}_i$ for short. The case with $m_i\neq \bar{m}_i$ does not bring substantial information and makes the expressions slightly more complicated.}
\begin{equation}
\alpha_{i} = b(j_i + 1) +\frac{1}{2b} \ , \ \ \ \ \ b^2=\frac{1}{k-2}. \label{alfabe}
\end{equation}

Formula (\ref{Ribault}) has been proven in Ref. \cite{R} for the case $|\Delta \omega |<n-2$; the case $|\Delta \omega |=n-2$, however, remained therein as a conjecture. In such case, the formula takes the suscinct form
\begin{eqnarray}
\left\langle \prod_{i=1}^{n} V_{j_i,m_i,\bar{m}_i}^{\omega_i } (z_i)\right>_{\text{WZW}} = {2\pi^{3-2n}b c^{2-n}} \prod_{i=1}^{n} \frac{\Gamma(-j_i-m_i)}{\Gamma(j_i+1+\bar{m}_i)}  \prod_{i<j} |z_i - z_j|^{2\beta_{ik}} \left\langle \prod_{i=1}^{n} V_{\alpha_i} (z_i) \right>_{\text{L}} .   \label{ghjghj310}
\end{eqnarray}

This formula for $|\Delta \omega |=n-2$, which unlike $|\Delta \omega |<n-2$ can not be proven using BPZ equation because of the absence of degenerate fields on the right hand side of (\ref{ghjghj310}), has been reproduced in Ref. \cite{Gaston} by means of the Coulomb gas formalism described in the previous section. We will review this derivation in the next section, and this will be the first step to construct our method to compute winding violating amplitudes. 

\section{Tree-level string amplitudes in AdS$_{3}$}

String amplitudes are defined by integrating correlation functions over the insertion of the vertices (the moduli) modulo the conformal Killing group (CKG) symmetries. At tree-level, one first considers the WZW correlation function (\ref{setitfri}) defined on the Riemann sphere, and then defines the string scattering
amplitudes as
\begin{equation}
\mathcal{A}=\int
\prod_{l=3}^{n-1}d^{2}z_{l}\ X_{n}^{\Delta \omega } \ \times \ ...
\label{stringtheory}
\end{equation}%
where the ellipses stand for the contribution of the internal CFT to the correlation functions. The integral in (\ref{stringtheory}) is over $n-3$ vertex insertions on $\mathbb{CP}^1$ in order to cancel out the volume of the CKG, $PSL(2,\mathbb{C})$. Without loss of generality, we can set $z_1=\infty $, $z_2=1$, and $z_n=0$. 

As explained before, correlation functions $X_{n}^{\Delta \omega }$ can be defined by resorting to $p$ operators of the representation (\ref{Vertex}) and $n-p$ operators of the representation (\ref{conjugated}). In the next subsections we will analyze several particular cases of (\ref{LA249}), which correspond to different values of $p$ and, consequently, to different values of the total winding number.




\subsection{Maximally winding violating amplitudes}

As a first example, let us review the computation of the maximally winding violating amplitude, carried out in Ref. \cite{Gaston}. Nevertheless, in order to avoid redundancies, and with the purpose to prepare the ingredients for the discussion of the genus-1 case, here we will compute this observable in an alternative way. In \cite{Gaston}, the amplitudes of a scattering process of $n$ strings in which the winding
number conservation is violated in $n-2$ was represented by the correlator
\begin{equation}
X_{n}^{2-n}=\left\langle  V
_{-1-j_{1},m_{1},\bar{m}_{1}}^{\omega
_{1}}(z_{1})\prod_{l=2}^{n}\tilde{V }_{j_{l},m_{l},%
\bar{m}_{l}}^{\omega _{l}}(z_{l}) \right\rangle_{\text{WZW}} . \label{laX}
\end{equation}


Working out the product of
the multiplicity factor coming from the Wick contraction of the $\beta $-$%
\gamma $ contribution, and resorting to the Dotsenko-Fateev integral representation \cite{DF1, DF2, DF3}, expression (\ref{laX}) was shown to realize (\ref{Ribault}) exactly. Here, instead of starting from (\ref{laX}), we will prefer to compute the same observable but using the alternative representation
\begin{equation}
X_{n}^{2-n}=\left\langle  V
_{-1-j_{1},m_{1},\bar{m}_{1}}^{\omega
_{1}}(z_{1})\prod_{l=2}^{n-1}\tilde{V }_{j_{l},m_{l},%
\bar{m}_{l}}^{\omega _{l}}(z_{l}){V }_{-1-j_{n},m_{n},%
\bar{m}_{n}}^{\omega _{n}}(z_n) \right\rangle_{\text{WZW}} . \label{laY}
\end{equation}

As discussed in subsection 2.5, the result has to be the same. However, as already said, prescription (\ref{laY}) is better suited for a path integral realization. 

Expanding (\ref{laY}), one obtains
\begin{eqnarray*}
X_{n}^{2-n} &=&{(-1)^{s}\Gamma (-s)c^{2-n}%
}
\prod_{l=2}^{n-1}\frac{(-1)^{-j_{l}-m_{l}}\Gamma (-j_{l}-\bar{%
m}_{l})}{\Gamma (1+j_{l}+m_{l})} \times \\
&&\times \int \prod_{r=1}^{s}d^{2}y_{r}\left\langle \gamma^{-1-j_{1}-m_{1}}(z_{1})\prod_{l=2}^{n-1}\beta^{j_{l}+m_{l}}(z_{l})\prod_{r=1}^{s}\beta(y_{r})
\gamma^{-1-j_{n}-m_{n}}(z_{\infty }) \right\rangle _{M=0} \times
\end{eqnarray*}%
\begin{equation*}
\times \left\langle e^{-\sqrt{\frac{2}{k-2}}(j_{1}+1)\phi
(z_{1})}\prod_{l=2}^{n-1}e^{-\sqrt{\frac{2}{k-2}}(j_{l}+\frac{k}{2}%
)\phi (z_{l})}\prod_{r=1}^{s}e^{-\sqrt{\frac{2}{k-2}}\phi
(y_{r})} e^{-\sqrt{\frac{2}{k-2}}(j_{n}+1)\phi
(\infty )} \right\rangle _{M=0} \times
\end{equation*}%
\begin{equation}
\times \left\langle e^{i\sqrt{\frac{2}{k}}m_{1}X(z_{1})}\prod_{l=2}^{n-1}e^{i\sqrt{\frac{2}{k}}(m_{l}-\frac{k}{2})X(z_{l})}
e^{i\sqrt{\frac{2}{k}}m_{n}X(z_{n})}
\right\rangle \left\langle \prod_{l=1}^{n}e^{i\sqrt{\frac{2}{k}}(m_{l}+%
\frac{k}{2}\omega _{l})T(z_{l})}\right\rangle \times h.c. \label{la47}
\end{equation}%
where $s$, in terms of the variables (\ref{alfabe}), is given by 
\begin{equation}
s=b^{-1}\sum\nolimits_{i=1}^{n}\alpha _{i}+1+b^{-2}. 
\end{equation}

The Wick contractions of the free fields yield the integral expression\footnote{Here, we have absorbed an irrelevant factor.}
\begin{equation*}
X_{n}^{2-n}=g_s^{2s}c^{2-n}\prod_{i=1}^{n}\frac{\Gamma (-j_{i}-\bar{m}%
_{i})}{\Gamma (1+j_{i}+m_{i})}\prod_{i<j}^{n}|z_{i}-z_{j}|^{2\beta _{ij}}
\times \end{equation*}%
\begin{equation}
\times \prod_{i<j}^{n}|z_{i}-z_{j}|^{-2\alpha _{i}\alpha _{j}}\Gamma
(-s)\int
\prod_{r=1}^{s}d^{2}y_{r}\prod_{i=1}^{n}\prod_{r=1}^{s}|z_{i}-y_{r}|^{-2\alpha
_{i}b}\prod_{r<t}^{s-1,s}|y_{r}-y_{t}|^{-2b^{2}},  \label{Uh}
\end{equation}%
with $\beta _{ij}=k/2-m_{i}-m_{j}-k\omega _{i}\omega
_{j}/2-m_{i}\omega _{j}-m_{j}\omega _{i}$. To obtain this, one also has to consider the combinatoric factor coming from the Wick contraction of the $\beta $-$\gamma $ system, now involving two operators with $\gamma$ fields inserted at $z_1$ and $z_n$. This factor can be seen to be
\begin{equation}
(-1)^s\frac{\Gamma(-j_1-\bar{m}_1)}{\Gamma(j_1+m_1+1)}\frac{\Gamma(-j_n-\bar{m}_n)}{\Gamma(j_n+m_n+1)}.
\end{equation} 

This expression yields the same integral representation as (\ref{laX}), cf. formula (24) of Ref. \cite{Gaston}. Therefore, the final results turns out to be the same as in \cite{Gaston}; namely
\begin{equation}
X_{n}^{2-n}=g_s^{2s}c^{2-n}\prod_{i=1}^{n}\frac{\Gamma (-j_{i}-\bar{m}%
_{i})}{\Gamma (1+j_{i}+m_{i})}\prod_{i<j}^{n}|z_{i}-z_{j}|^{2\beta _{ij}}
\left\langle \prod_{i=1}^{n}V_{\alpha _{i}}(z_{i})\right\rangle _{%
\text{L}},  \label{final}
\end{equation}%
which coincides with the formula conjectured in Ref. \cite{R} for the maximally violating correlator. This shows with a particular working example how the prescription involving two operators in the representation (\ref{Vertex}) and $n-2$ operators in the representation (\ref{conjugated}) also leads to the correct result on the sphere topology, at least in the case of maximally winding violating correlators, $\Delta \omega =2-n$. Now, let us move to the case of next-to-maximally winding violating correlators.

\subsection{Next-to-maximally winding violating amplitudes}

The maximally winding violating correlators discussed above correspond to the case $p=1$ in (\ref{LA249}). Now, let us consider the
case $p=2$, which describes next-to-maximally winding violating amplitudes,
i.e. with $\Delta \omega =3-n$. According to the prescription described before, this would be given by the correlator 
\begin{equation}
X_{n}^{3-n}=\left\langle 
V _{-1-j_{1},m_{1},\bar{m}_{1}}^{\omega
_{1}}(z_{1})\prod_{l=2}^{n-1}\tilde{V }_{j_{l},m_{l},%
\bar{m}_{l}}^{\omega _{l}}(z_{l}) V_{-1-j_n,m_n,\bar{m}_n}^{\omega_n} \right\rangle_{\text{WZW}} ,  \label{hjhjk}
\end{equation}%
which yields
\begin{eqnarray*}
X_{n}^{3-n} &=&(-1)^sg_s^{2s}\Gamma(-s)c^{3-n}\prod_{l=2}^{n-1}\frac{(-1)^{-j_{l}-m_{l}}%
\Gamma (-j_{l}-\bar{m}_{l})}{\Gamma (1+j_{l}+m_{l})} \\
&&\int \prod_{r=1}^{s}d^{2}y_{r}\left\langle \gamma
^{-1-j_{1}-m_{1}}(z_{1})\prod_{l=2}^{n-1}\beta
^{j_{l}+m_{l}}(z_{l})\prod_{r=1}^{s}\beta (y_{r}) \gamma
^{-1-j_{n}-m_{n}}(z_{n})\right\rangle_{M=0} 
\end{eqnarray*}%
\begin{equation*}
\left\langle e^{-\sqrt{\frac{2}{k-2}}(j_{1}+1)\phi (z_{1})}\prod_{l=2}^{n-1}e^{-\sqrt{\frac{2}{k-2}%
}(j_{l}+\frac{k}{2})\phi (z_{l})}\prod_{r=1}^{s}e^{-\sqrt{\frac{2}{%
k-2}}\phi (y_{r})} e^{-\sqrt{\frac{2%
}{k-2}}(j_{n}+1)\phi (z_{n})}\right\rangle _{M=0}
\end{equation*}%
\begin{eqnarray}
&&\left\langle e^{i\sqrt{\frac{2}{k}}m_{1}X(z_{1})}\prod_{l=2}^{n-1}e^{i\sqrt{\frac{2}{k}}(m_{l}-\frac{k}{%
2})X(z_{l})} e^{i\sqrt{\frac{2}{k}}%
m_{n}X(z_{n})} \right\rangle  \left\langle \prod_{l=1}^{n}e^{i\sqrt{\frac{2}{k}}(m_{l}+\frac{k}{%
2}\omega _{l})T(z_{l})}\right\rangle \times h.c.  \label{wwwww}
\end{eqnarray}%
with 
\begin{equation}
s=-2-\sum_{l=1}^{n}j_l-\frac{k}{2} (n-3) , \ \ \ \sum_{i=1}^{n}m_i=\sum_{i=1}^{n}\bar{m}_i = \frac{k}{2} (n-3), \ \ \ \sum_{i=1}^{n}\omega_i = 3-n.
\end{equation}

After Wick contraction of free field propagators, the integral representation of the correlator (\ref{wwwww}) yields
\begin{equation}
\Gamma(-s) (-1)^s \prod_{i=1}^n \frac{\Gamma(-j_i-m_i)}{\Gamma(j_i+m_i+1)} \prod_{i<t}^{n} |z_i - z_t|^{2\beta_{ij}-4\alpha_i \alpha_j } \int \prod_{r=1}^{s} d^2w_r \prod_{r=1}^{s}\prod_{i=1}^{n} |z_i - w_t|^{-4b\alpha_i } \prod_{r<t}^{s} |w_r - w_t|^{-4b^2}, \label{345}
\end{equation}
where, again, we have used the Wick contraction of the ghost system and the variables (\ref{alfabe}) to express the final result. 

However, notice that (\ref{345}) is not exactly what one would naively expect from the $H_3^+$ WZW-Liouville correspondence formula (\ref{Ribault}). In addition, (\ref{Ribault}) includes the insertion of one degenerate operator $V_{-\frac{1}{2b}}(x)$ that seems to be missing in (\ref{345}). Such insertion would generate an additional factor
\begin{eqnarray}
\mathcal{I} =\int
d^{2}x\prod_{r=1}^{s}|x-w_{r}|^{2}\prod%
_{i=1}^{n}|z_{i}-x|^{2j_{i}+2m_{i}} . \label{I} 
\end{eqnarray}%
This does not mean that formula (\ref{345}) is incorrect; in fact, the case $n=3$ have been explicitly computed in \cite{Becker, GN3} and shown to reproduce, through a Melian transform, the correct result \cite{Teschner1, Teschner2}. What the lack of the extra insertion in $x$ actually means is that the form (\ref{345}) does not suffice to make the correspondence between $H_3^+$ WZW and Liouville correlators manifest. In order to solve this inconvenient, one possible strategy would be trying to solve the Selberg type integral (\ref{I}) by standard techniques and see whether it produces the right dependence on $z_i$ ($i=1,2,...n$) and $w_r$ ($r=1,2,...s$) needed to reproduce (\ref{345}). However, this is not efficient; there is a more proactive strategy, which is trying to interpret the Liouville degenerate insertions $V_{-\frac{1}{2b}}(x)$ from the AdS$_3$ worldsheet point of view. We will see that, in fact, the Liouville degenerate fields can be thought of as part of a dimension-$(1,1)$ field in the WZW theory\footnote{A similar idea, in a different representation, has been considered in Ref. \cite{Yo}.}. This will allow us to give a general prescription for computing winding violating correlators and eventually ``exponentiate'' such operator. Let us see how it works: The first observation is that, in (\ref{Ribault}), the Liouville degenerate vertex comes with an accompanying factor 
\begin{equation}
\prod_{i=1}^{n}|z_{i}-x|^{-k+2m_{i}}V_{-\frac{1}{2b}}(x) . \label{ooo}
\end{equation}%
Besides, the OPE\ of the degenerate operator $V_{-\frac{1}{2b}}(x)$ and the other vertices gives%
\begin{eqnarray}
V_{-\frac{1}{2b}}(x)V_{\alpha _{i}}(z_{i}) &=&e^{-\frac{1}{\sqrt{2}b}\varphi
(x)}e^{\sqrt{2}\alpha _{i}\varphi (z_{i})}\simeq |z_{i}-x|^{\frac{2\alpha
_{i}}{b}}, \label{518}
\end{eqnarray}%
which, in particular, yields the following OPE with the Liouville screening operator 
\begin{eqnarray}
V_{-\frac{1}{2b}}(x)V_{b}(y_{r}) &=&e^{-\frac{1}{\sqrt{2}b}\varphi (x)}e^{%
\sqrt{2}b\varphi (y_{r})}\simeq |x-y_{r}|^{2}. \label{519}
\end{eqnarray}%

These $x$-dependent factors indicate how to interpret the insertion of an operator at $x$ in the WZW theory side. If the original question was how to realize such a contribution from the string
worldsheet point of view, the answer turns out to be that (\ref{I}) can be realized by
the introduction of the dimension-$(1,1)$ operator
\begin{equation}
\tilde{V }_{1-k,k,k}^{-2}(x)=\frac{1}{c} \beta (x)e^{\sqrt{\frac{k-2}{2}}\phi (x)}e^{i\sqrt{\frac{k}{2}%
}X(x)}\times h.c.  \label{TheGuy}
\end{equation}%
which, besides, exhibits the regular OPE
\begin{equation}
J^3(z)\tilde{V }_{1-k,k,k}^{-2}(x)\simeq {\mathcal O}(1).
\end{equation}

This is quite interesting as it permits to undesrand the presence of the Liouville fields (\ref{degenerado}) from the string worldsheet point of view as a kind of a marginal deformation. Furthermore, the insertion of operator (\ref{TheGuy}) happens to be consistent with the adequate charge conjugation conditions
\begin{equation}
N_{\beta }(X_{n}^{3-n})-N_{\gamma }(X_{n}^{3-n})= N_{\bar{\beta } }(X_{n}^{3-n})-N_{\bar{\gamma } }(X_{n}^{3-n})=1,
\end{equation}
and with the total winding number
\begin{equation}
\sum_{i=1}^{n} \omega_i = 3-n. \label{522}
\end{equation}

In conclusion, the WZW correlator to be considered to describe a next-to-maximally winding violating process is
\begin{equation}
X_{n}^{3-n}=\int d^2x \ \left\langle 
V _{-1-j_{1},m_{1},\bar{m}_{1}}^{\omega
_{1}}(z_{1}) \tilde{V}_{1-k,k,k}^{-2}(x) \prod_{l=2}^{n-1}\tilde{V }_{j_{l},m_{l},%
\bar{m}_{l}}^{\omega _{l}}(z_{l}) V_{-1-j_n,m_n\bar{m}_n}^{\omega_n }(z_n) \right\rangle_{\text{WZW}} , \label{524}
\end{equation}
cf. (\ref{hjhjk}). Taking into account (\ref{ooo})-(\ref{519}), it is easy to verify that correlator (\ref{524}) actually realizes the $H_3^+$ WZW-Liouville correspondence formula (\ref{Ribault}) in the case $\Delta \omega =3-n$. This already suggests a prescription to compute winding violating amplitudes, which actually refines the formalism of \cite{GN2}. In this new version, instead of inducing the violation of the winding number by playing with the relative amount of operators (\ref{Vertex}) and operators (\ref{conjugated}) in the correlator as in \cite{GN2}, we are obtaining the same effect by inserting dimension-$(1,1)$ operators in the worldsheet. Roughly speaking, the role played by operator $\tilde{V}_{1-k,k,k}^{-2}(x)$ in (\ref{524}) is that of reducing the winding number violation in one unit, starting from the maximally violating case (\ref{laY}). In some sense, this method is the inverse of the Fateev-Zamolodchikov-Zamolodchikov prescription, reviewed by Maldacena and Ooguri in \cite{MO3}, in which the insertion of the dimension-$(0,0)$ ``spectral flow operator'' plays the role of increasing the violation of the winding number in one unit.

\subsection{Next-to-next-to-maximally winding violating amplitudes}

Before going to the general case, let us test our proposal and gain intuition by seeing how it works in another special case: the next-to-next-to maximally winding violating amplitude. According to the new prescription we have just discussed, this would correspond to the correlator 
\begin{eqnarray}
X_{n}^{4-n}&=&\int d^2x_1 \int d^2x_2  \left< 
V _{-1-j_{1},m_{1},\bar{m}_{1}}^{\omega _{1}}(z_{1}) \right. \tilde{V }_{1-k,k,k}^{-2}(x_1) \tilde{V }_{1-k,k,k}^{-2}(x_2)\prod_{l=2}^{n-1}\tilde{%
V }_{j_{l},m_{l},\bar{m}_{l}}^{\omega _{l}}(z_{l})  \times \nonumber \\
&& \left. \times V _{-1-j_{n},m_{n},\bar{n}_{n}}^{\omega _{n}}(z_{n})
\right\rangle _{\text{WZW}} ,  \label{loquequiero}
\end{eqnarray}%
where two operators (\ref{TheGuy}) have been introduced. The charge compensation conditions in this case read
\begin{equation}
(k-2)-\sum_{i=1}^{n}j_i-\frac{k}{2} (n-2)-s=1 \ , \label{ryt}
\end{equation}
together with
\begin{equation}
\sum_{i=1}^{n}m_i = \sum_{i=1}^{n}\bar{m}_i = \frac{k}{2} (n-4) \ ,   
\end{equation}
and
\begin{equation}
\sum_{i=1}^{n}\omega _i = 4-n. 
\end{equation}
Using the variables (\ref{alfabe}), condition (\ref{ryt}) can be written as
\begin{equation}
\sum_{i=1}^{n}\alpha_i  -\frac{1}{b} +bs = Q, 
\end{equation}
which exactly matches the Liouville relation (\ref{pupo2}) for $n+2$ operators with $\alpha_{n+1}=\alpha_{n+2}=-1/(2b)$. In addition, one observes that
\begin{equation*}
\tilde{V }_{1-k,k,k}^{-2}(x_{1})\tilde{V }_{1-k,k,k}^{-2}(x_{2})\ \simeq
|x_{1}-x_{2}|^{2},
\end{equation*}%
which exactly reproduces the contribution%
\begin{equation}
|x_{1}-x_{2}|^{k}V_{-\frac{1}{2b}}(x_{1})V_{-\frac{1}{2b}}(x_{2})\simeq
|x_{1}-x_{2}|^{2}, \label{El2}
\end{equation}%
also present in the formula (\ref{Ribault}). Therefore, also in the case $|\Delta \omega |<n-3$, the correct expression for the WZW correlators on the sphere is obtained from the Coulomb gas realization
\begin{equation}
\int \prod_{a=1}^{d} d^2x_a
\int \prod_{r=1}^{s} d^2w_r \left\langle
\prod_{l=1}^{2} {V }_{-1-j_j,m_l,\bar{m}_l}^{\omega _l}(z_l)
\prod_{i=3}^{n} \tilde{V }_{j_i,m_i,\bar{m}_i}^{\omega_i}(z_i)
\prod_{a=1}^{d} \tilde{V }_{1-k,k,k}^{-2}(x_a)
\prod_{r=1}^{s} \tilde{V }_{1-\frac{k}{2},\frac{k}{2},\frac{k}{2}}^{-1}(w_r)
\right\rangle_{M=0}
\label{LaK}
\end{equation}
with $d=n-2-|\Delta\omega |$ and $s= \sum_{i=1}^{n}j_i+1+(k/2)(n-2+d)$.

This confirms that the presence of the dimension-$(1,1)$ operator $\tilde{V }_{1-k,k,k}^{-2}(x)$ in the correlator controls the winding number conservation. The question whether the introduction of a dimension-$(1,1)$ operator could be behind the violation of the winding number had already been raised in Ref. \cite{HM}, where a no-go argument for the existence of such operator was presented. The argument therein was basically that if such operator existed then this would lead to a violation of the winding number in an arbitrary amount which in principle could violate the bound $|\Delta \omega |\leq n-2$. The reason why  here we managed to circumvent this obstruction is the contribution (\ref{El2}) in the integrand when computing the correlator. As explained in \cite{FH}, the accumulation of factors $|x_a-x_b|^2$ in the integrand of the Dotsenko-Fateev type formula can eventually yield a vanishing result\footnote{See subsection 5.4 in comparison with the analysis carried out in Ref. \cite{FH}.}. This is what happens, for instance, in sine-Liouville theory, where the violation of the winding number is also bounded from above although it is precisely produced by the screening operators of the theory \cite{FH}.

Based on (\ref{LaK}), we have all the ingredients to state the general prescription. However, we will postpone this to subsection 5.2, where we will present a more general case that will be valid both for genus-zero ($g=0$) and for genus-one ($g=1$) correlation functions

\section{Genus-one amplitudes in AdS$_3$}

\subsection{On winding number violation at genus-$g$}

Before going into the discussion of the Coulomb gas representation of genus-1 correlators, let us make a digression about the bound on the violation of the winding number on a genus-$g$ surface. The argument follows from factorization of the $g$-loop process or, in other words, from the seewing of the genus-$g$ Riemann surface: An $n$-string scattering amplitudes at $g$ loops is associated to a genus-$g$ $n$-puncture surface, which can always be decomposed in $3g-3+n$ tubes and $2g-2+n$ trinions. In principle, as it happens in the sphere, in a genus-$g$ amplitude with $n$ external states the total winding number can be violated; namely, there are in principle non-vanishing amplitudes satisfying
\begin{equation}
\Delta \omega \equiv \sum_{e=1}^{n}\omega_e  \neq 0 . \label{ww}
\end{equation}

Then, the question is whether there is a natural upper bound for $|\Delta \omega |$ as a funcion of $n$ and $g$. Factorizing the $n$-string amplitude, and taking into account that for $g=0$, $n=3$ one has $|\Delta \omega |\leq 1$, one finds the bound $|\Delta \omega |\leq n-2+2g$. To see this explicitly, consider
\begin{equation}
\Delta \omega = \sum_{e=1}^{n}\omega_e = \sum_{V=1}^{2g-2+n} \sum_{l=1}^{3} \omega_l^{(V)}  \label{suma}
\end{equation}
where the sum with index $V$ in (\ref{suma}) runs over the $2g-2+n$ vertices (trinions) of the diagram, with $\omega^{(V)}_l$ being the winding number of the $l^{\text{th}}$ state $(l=1,2,3)$ converging in the $V^{\text{th}}$ trinion. This yields
\begin{equation}
\left\vert \Delta \omega \right\vert  \leq \sum_{V=1}^{2g-2+n}  \left\vert  \sum_{l=1}^{3} \omega^{(V)}_l \right\vert \leq 2g-2+n , \label{tyu65}
\end{equation}
where we have used that on each trinion, i.e. on each $3$-puncture sphere, $|\Delta \omega |\leq 1$. In particular, we conclude that the natural bound for $|\Delta \omega |$ in a 1-loop $n$-string amplitudes is $|\Delta \omega | \leq n$. This is consistent with the result of \cite{HikidaSchomerus}, which shows that the genus-$g$ WZW correlation functions that preserve the winding are described by $(2n-2+2g)$-point Liouville correlators with $n$ degenerate insertions. It is natural to especulate that, as it happens with the sphere, removing a denerate field from the dual Liouville genus-$g$ correlator would result in increasing the winding number violation in 1 unit, and in principle one can iterate this procedure {\it ad nutum}. Nevertheless, it is worth pointing out that upper bound (\ref{tyu65}) is kinematical and does not mean that there are no stricter dynamical bounds in the theory. For instance, one expects the winding number to be conserved in the 2-point functions.

\subsection{A general prescription}

The $n$-string scattering amplitudes in AdS$_3$ at $g$ loops is given by $n$-point correlation functions of the $SL(2,\mathbb{R})$ WZW theory on the genus-$g$ surface. As discussed in section 4, in a genus-$g$ $n$-point function (with $g=0,1$) that involves $n_+$ fields $\tilde{V }^{\omega_i}_{j_i, m_i, \bar{m}_i} (z_i) $ (with $i=1,2,...n_+$), $n_-=n-n_+$ fields $V^{\omega_l}_{-1-j_l, m_l, \bar{m}_l} (z_l) $ (with $l=n_+ +1,n_+ +2,...n$), and $d$ fields $\tilde{V }_{1-k,k,k}^{-2}(x_a) $ (with $a=1,2,...d$), the amount of screening operators, $s$, is given by
\begin{equation}
-\sum_{i=1}^{n_-} (j_i+1) -\sum_{l=1}^{n_+} (j_l +{k}/{2}) -s + d ({k}/{2}-1) = -(1-g), \label{harrib}
\end{equation}
which, written this in terms of Liouville variables (\ref{alfabe}), reads
\begin{equation}
\sum_{i=1}^n \alpha_i + bs -\frac{1}{2b} (2g-2+d+n_{-}) = (1-g) Q.
\end{equation}

The integration over the zero-mode of the fields $X(z)$ and $T(z)$, on the other hand, yields 
\begin{equation}
\sum_{i=1}^n m_i = \sum_{i=1}^n \bar{m}_i = \frac{k}{2} (n_{+} -d) \ ,  \ \ \ \sum_{i=1}^n \omega_i = d-n_{+} .
\end{equation}

Making contact with the formula (\ref{Ribault}) and its genus-1 generalization of Ref. \cite{HikidaSchomerus}, and demanding the amount of degenerate fields in the Liouville correlator to match the amount of operators $\tilde{V }_{1-k,k,k}^{-2}(x)$ in the WZW correlator, one finds
\begin{equation}
n_{-}=2-2g.
\end{equation}

This means that the WZW correlators have to be defined with $n_{+}=n-2$ and $n_{-}=2$ at genus $g=0$, and with $n_{+}=n$ and $n_{-}=0$ at genus $g=1$. Then, at genus $g=0$ the prescription is (\ref{LaK}), which yields
\begin{eqnarray}
X^{\Delta \omega }_n &=& \frac{c^{-\Delta \omega }}{\Gamma (d+1)} \prod_{i=1}^{n} \frac{\Gamma (-j_i -\bar{m}_i) }{\Gamma (j_i+1+m_i)} \prod_{i<j}^{n} |z_i -z_j |^{2\beta_{ij}}  \int \prod_{a=1}^{d} d^2x_a 
\prod_{a<b}^{d} |x_a -x_b |^{k} 
\prod_{a=1}^{d} \prod_{i=1}^{n}|x_a -z_i |^{2m_i-k} \times \nonumber \\
&& \left\langle  \prod_{i=1}^{n} V_{\alpha_i }(z_i) \prod_{a=1}^{d} V_{-\frac{1}{2b}}(x_a) \right\rangle _{\text{L}}
\end{eqnarray} 
with $\beta_{ij}=k-2m_i-2m_j-k\omega_i \omega_j -2m_i \omega_j -2m_j \omega_i $, $d=n-2-|\Delta \omega |$, $b^{-2}=k-2$, $\alpha_i = b(j_i+k/2)$. By construction, this exactly matches (\ref{Ribault}). At genus-$1$, on the other hand, the $n$-point function involves $n$ fields $\tilde{V }^{\omega_i}_{j_i, m_i, \bar{m}_i} (z_i) $ and $d=n-|\Delta \omega |$ fields $\tilde{V }_{1-k,k,k}^{-2}(x_a)$, and it is given by the correlator\footnote{For short, we omit the $b-c$ ghost contribution and other factors that are not relevant for our discussion.}
\begin{equation}
\int \prod_{a=1}^{n-|\Delta \omega |} d^2x_a \int \prod_{r=1}^{s} d^2w_r
\left\langle  
\prod_{i=1}^{n} \tilde{V }^{\omega_i }_{j_i , m_i , \bar{m}_i } (z_i ) 
\prod_{a=1}^{n-|\Delta \omega |} \tilde{V }^{-2}_{1-k,k,k} (x_a, )
\prod_{r=1}^{s} \tilde{V }^{-1}_{1-\frac{k}{2},\frac{k}{2},\frac{k}{2}} (y_r) 
\right\rangle _{M=0} , \label{D620}
\end{equation} 
with $s=-\sum_{i=1}^{n} j_i-|\Delta \omega |(k-2)/2 -n$.


In the next subsections, we will discuss the free field realization of (\ref{D620}) on the torus.

\subsection{The theory on the torus}

In order to extend out Coulomb gas formalism to the case of 1-loop amplitudes it is necessary to formulate the free field theory on the genus-1 surface: Consider a genus-1 surface of modular parameter $\tau =\tau _1 + i \tau _2$, with $\tau_{1,2}\in \mathbb{R}$. This is defined as $\mathbb{C}/\Gamma $, with $\Gamma $ being the lattice generated by the operations
\begin{equation}
z \to z+1 \ , \ \ \ \ z\to z+\tau .  \label{rt61}
\end{equation}

When trying to define the theory (\ref{sigma}) on the torus, one could be tempted to impose the invariance of the fields under (\ref{rt61}); however, this is too restrictive. There are twisted sectors in the theory that, while leaving (\ref{sigma}) invariant, are quasi-periodic in the fields \cite{Gawedzky}. These sectors are defined by imposing the following conditions for the fields on the lattice
\begin{eqnarray}
\beta (z+p+q\tau ) = e^{2\pi iq\lambda } \beta (z) \ , \ \ \ \gamma (z+p+q\tau ) = e^{-2\pi iq\lambda } \gamma (z)  ,  \label{QA} 
\end{eqnarray}
together with
\begin{eqnarray}
\phi (z+p+q\tau ) = \phi (z) - 2\pi q\sqrt{2(k-2)} \text{Im}(\lambda ) , \label{QC}
\end{eqnarray}
with $(q,p)\in \mathbb{Z}^2$ and $\lambda \in \mathbb{C}$ being an arbitrary variable with real part $\text{Re}(\lambda )= (\lambda + \bar{\lambda})/2$ and imaginary part $\text{Im}(\lambda )= (\lambda - \bar{\lambda})/(2i)$. Different values of $\lambda $ describe different twisted sectors of the theory.

On the torus, there is a unique non-constant classical solution of the equations of motion satisfying the above boundary conditions \cite{Gawedzky, HikidaSchomerus}. This corresponds to $\beta _0 = \gamma _0 =0$ and $\phi_{0}  (z) = 2\pi \sqrt{2(k-2)} {\text{Im}(z)\text{Im}(\lambda)}/{\text{Im}(\tau)}$. Then, it is convenient to decompose the field $\phi (z)$ in its classical part $\phi_{0} (z)$ and its fluctuations $\phi (z) - \phi_{0} (z)$. While the correlators will be computed in terms of operators evaluated on the fluctuations, an overall factor gathering the exponential fields evaluated on the classical solution could appear; see \cite{HikidaSchomerus}. 

As we will see, the relevant contribution in the genus-$g$ correlators will come from the exponential part of the vertex operators. Then, the main ingredient we need is the Green function on the torus. This is given by
\begin{equation}
\partial \bar{\partial } \left\langle \phi (z) \phi (w) \right\rangle = -2\pi \delta^{(2)}(z-w)+\frac{4\pi }{\text{Im}(\tau)} , \label{khgjkhgfds}
\end{equation}
where the $\delta^{(2)}(z)$ is the periodic Dirac $\delta $-function on $\mathbb{C}$ with the required identifications; namely 
\begin{equation}
\delta^{(2)}(z) = \sum_{(p,q)\in\mathbb{Z}^2} \delta^{(2)}(z-p-q\tau )  .
\end{equation}

The second term on the right hand side of (\ref{khgjkhgfds}) is the background charge. It stands in the Poisson equation to cancel the charge of the single particle on the compact 
surface; it corresponds to a charge density uniformly distributed over the volume of the torus, $\text{Im}(\tau)$.

The solution to (\ref{khgjkhgfds}) is given by
\begin{equation}
\left\langle \phi (z) \phi (w) \right\rangle = -2\log  \left\vert \frac{\theta_{1} (z-w|\tau ) }{\partial \theta_{1}  (0|\tau )} e^{-\pi \frac{(\text{Im}(z-w))^2}{\text{Im}(\tau )}} 
\right\vert \equiv -2\log \chi(z-w|\tau ) \ ,  \label{Ula515}
\end{equation}
where $\theta_{1} $ is the Jacobi function
\begin{equation}
\theta_{1} (z|\tau ) = - \sum_{n\in \mathbb{Z}} e^{i\pi (n+\frac{1}{2})^2 \tau +2\pi i (n+\frac{1}{2}) (z+\frac{1}{2}) } ,
\end{equation}
which satisfies
\begin{equation}
\theta_{1} (z+\tau |\tau ) = - e^{- \pi i \tau-2\pi i z} \theta_{1} (z|\tau ) \ , \ \ \ \theta_{1} (z+1 |\tau ) = - \theta_{1} (z|\tau ) .
\end{equation}

From this, it is possible to verify that function
\begin{equation}
\chi (z|\tau ) = e^{-\frac{\pi (\text{Im} (z))^2}{\text{Im}(\tau )}}  \left\vert  \frac{\theta_{1} (z|\tau )}{\partial \theta_{1}  (0|\tau )}  \right\vert , \label{hgh}
\end{equation}
satisfies the required periodicity conditions
\begin{equation}
\chi (z+p+q\tau |\tau )  = \chi (z|\tau ) .
\end{equation}
and has the appropriate short distance behavior, yielding a logarithmic dependence when $z\simeq 0$.

With correlator (\ref{Ula515}), we can expand (\ref{D620}) and write\footnote{Here, we are omitting irrelevant factors.}
\begin{eqnarray}
X^{\Delta \omega }_n &=& \frac{c^{-\Delta \omega }}{\Gamma (d+1)} \prod_{i=2g+1}^{n} \frac{\Gamma (-j_i -\bar{m}_i) }{\Gamma (j_i+1+m_i)} \int \prod_{a=1}^{d} d^2x_a 
\prod_{a<b}^{d} \chi (x_a -x_b |\tau )^{k}
\prod_{i<j}^{n} \chi (z_i -z_j |\tau )^{2\beta_{ij}} 
  \times \nonumber \\
&& \times \prod_{a=1}^{d} \prod_{i=1}^{n} \chi (x_a -z_i |\tau )^{2m_i-k}
\left\langle  \prod_{i=1}^{n} V_{\alpha_i }(z_i) \prod_{a=1}^{n-|\Delta \omega |} V_{-\frac{1}{2b}}(x_a) \right\rangle _{\text{L}} 
\end{eqnarray} 
with $d=n-|\Delta \omega |$. This follows from noticing that, at $g=1$
\begin{equation}
\left\langle  \prod_{i=1}^{n} e^{-\sqrt{\frac{2}{k-2}}(j_i+\frac{k}{2})\phi (z_i)}  \prod_{a=1}^{d} e^{\sqrt{\frac{k-2}{2}}\phi (x_a)} \right\rangle_{\text{WZW}} = \left\langle  \prod_{i=1}^{n} V_{\alpha_i }(z_i) \prod_{a=1}^{n-|\Delta \omega |} V_{-\frac{1}{2b}}(x_a) \right\rangle _{\text{L}} ,
\end{equation}


An overall factor $|\det \partial\bar{\partial }_{\lambda }|^{-1}$ also appears in (\ref{D620}). This comes from the $\beta $, $\bar{\beta }$ fields, and the subscript $\lambda $ refers to the determinant on twisted differentials. It has been shown in \cite{Gawedzky} that, surprisingly, this determinant is associated to the quotient between the genus-1 partition function of the WZW theory and that of a free field theory with a background charge. This has been used in \cite{HikidaSchomerus} to work out a genus-1 generalization of the $H_3^+$ WZW-Liouville correspondence and relate the corresponding partition functions.


The one-loop string amplitudes are then given by integrating the genus-1 correlation functions over the inserting points of the vertex operators. Formaly, one has 
\begin{equation}
{\mathcal A} = \frac{1}{Z} \int_{{\mathcal F}} \frac{d^2 \tau }{2\text{Im}(\tau)} \int \prod_{i=1}^{n}d^2z_i \ X_n^{\Delta \omega }\times \ ... \label{uuu}
\end{equation}
where the ellipses stand for the contribution of the internal CFT. In (\ref{uuu}), we are integrating over the $n$ vertex operators, what amounts to average over the inserting points of the $n^{\text{th}}$ operator in order to cancel out the volume of the CKG. On the torus, the latter corresponds to the translations, and then it generates the measure $\int d^2z_n(2\text{Im}(\tau))^{-1}$. In string theory, one typically integrates over the fundamental region ${\mathcal F}_0: (|\tau |>1,$ $\text{Re}(\tau )<1/2$, $\text{Im}(\tau )>0)$, in order to avoid redundancies under the $PSL(2,\mathbb{Z})$ modular transformations. In Ref. \cite{Gawedzky} the integration over the fundamental region was discussed in relation to the twisted sectors of the $H_3^+$ WZW model. In Ref. \cite{MO2}, the discussion of \cite{Gawedzky} was reconsidered taking into account the contribution of the spectrally flowed sectors, and the string theory one-loop partition function was shown to be modular invariant. We will not repeat these discussions here; instead, we refer to the Refs. \cite{Gawedzky, GN2} and Ref. \cite{HikidaSchomerus}.

\subsection{Exponentiation}

In conclusion, we have presented a Coulomb gas representation of $n$-point correlation functions in the $SL(2,\mathbb{R})$ WZW model which is suitable to compute three-level and one-loop scattering amplitudes of winding string states in AdS$_3$ space. This is a refined version of the free-field formalism of Ref. \cite{GN2}, which has shown to describe correctly the maximally winding violating processes \cite{GN3, Gaston}. Here, we have extended such formalism in two directions: On the one hand, we have explained how to compute the next-to-maximally winding violating process. On the other hand, he have proposed a prescription to extend the computation to the genus-1 Riemann surface. A central role in such prescription is played by the dimension-$(1,1)$ operator \cite{TheGuy}, which we have identified as the object that controls the winding number conservation. This follows from interpreting the degenerate fields that appear in the $H_3^+$ WZW-Liouville correspondence from the AdS$_3$ worldsheet point of view. That is, our prescription follows from writing WZW correlation functions as the convolution of Liouville field theory correlation functions. This is consistent with the exponentiation of the operator (\ref{TheGuy}), which therefore can be thought of as a perturbation of the action. Explicitly, the genus-$g$ correlation functions (with $g=0,1$) result to be given by 
\begin{equation}
\prod_{i=3-2g}^{n}\frac{c\Gamma (-j_i-\bar{m}_i)}{\Gamma (1+j_i+{m}_i)}
\left\langle  \exp \left( { \int \tilde{V }_{1-k,k,k}^{-2} (x)d^2x}\right)   \prod_{l=1}^{2-2g} V_{-1-j_l, m_l, \bar{m}_l }^{\omega_l } (z_l)
\prod_{i=3-2g}^{n} \tilde{V}_{j_i, m_i, \bar{m}_i }^{\omega_i } (z_i)   \right\rangle _{\text{WZW}} , \label{ultimaycerramo}
\end{equation}
where the different terms in the expansion of the perturbation describe process with different total winding number.

It is worth noticing that, in addition to operator $\tilde{V }_{1-k,k,k}^{-2} (x)$, in (\ref{ultimaycerramo}) one can consider its complex conjugate
\begin{equation}
\Lambda (x) = c^{-1/2} \beta (x) e^{\sqrt{\frac{k-2}{2}}\phi (x)} e^{-i\sqrt{\frac{k}{2}}X(x)} \times h.c. \label{Finaaaa}
\end{equation}
which is also a dimension-$(1,1)$ operator with regular OPE with the currents $J^3(z)$, $\bar{J}^3(\bar{z})$. In contrast to $\tilde{V }_{1-k,k,k}^{-2} (x)$, $\Lambda (x)$ does not belong neither to representation (\ref{Vertex}) nor (\ref{conjugated}). Nevertheless, being the complex conjugate to (\ref{TheGuy}), it can be added to the action in order to compose the real interaction term
\begin{equation}
\frac{2}{c} \int d^2x \ \beta \bar{\beta } \ e^{\sqrt{\frac{k-2}{2}} \phi} \cos \left( \sqrt{{k}/{2}} X \right) . \label{sineLiouville} 
\end{equation} 

This interaction term has evident resemblance to the interaction term in the sine-Liouville theory, which has been conjectured to be T-dual to the $SL(2,\mathbb{R})/U(1)$ WZW theory \cite{FZZ, HikidaSchomerus2}. It is possible to argue that this is more than a reminiscence: The presence of operator (\ref{sineLiouville}) in the action together with the possibility of neutralizing the contribution of the $\beta $-$\gamma $ system by resorting to the conjugate representation (\ref{conjugated}), leads to the realization of sine-Liouville correlators if one identify the exponential part of operators as $\tilde{V }_{j,m,\bar{m}}^{\omega }  \leftrightarrow \tilde{V }_{-j-\frac{k}{2},m-\frac{k}{2},\bar{m}-\frac{k}{2}}^{\omega +1}$. To see this explicitly, notice that in the expansion of the exponential of the interaction term (\ref{sineLiouville}), the contribution with $d_+$ operators (\ref{TheGuy}) and $d_-$ operators (\ref{Finaaaa}) yields the charge conservation conditions
\begin{equation}
\sum_{i=1}^{n} j_i +\frac{k}{2}n + s-\frac{(k-2)}{2}(d_+ + d_-)=1-g, 
\end{equation}
together with
\begin{equation}
\sum_{i=1}^{n} m_i  = \frac{k}{2} (n - d_+ + d_-) \ , \ \ \ \sum_{i=1}^{n} \omega_i  = n + d_+ - d_-  
\end{equation}
which exactly parallels the realization of sine-Liouville correlation functions \cite{KKK, FH} after making $(j,m,\bar{m},\omega )\to (-j-k/2,m-k/2,\bar{m}-k/2,\omega +1)$. It would be interesting to further explore this in connection with the Fateev-Zamolodchikov-Zamolodchikov conjecture, cf. \cite{Yo, HikidaSchomerus2}.

\[
\]

This paper is based on a series of talks that the author has delivered at Universiteit van Amsterdam (UvA), Universit\'e Libre de Bruxelles (ULB), and South-American Institute for Fundamental Research (SAIFR-ICTP). The author thanks those institutions for the hospitality. This work has been partially funded by FNRS-Belgium (convention FRFC PDR T.1025.14 and convention IISN 4.4503.15), by CONICET of Argentina, by the Communaut\'{e} Fran\c{c}aise de Belgique through the ARC program and by a donation from the Solvay family.


\begin{thebibliography}{99}

\bibitem{MO1} J.M. Maldacena and H. Ooguri, \textit{Strings in AdS}$_{%
\mathit{3}}$\textit{\ and the SL(2,R) WZW Model. Part 1: The Spectrum}, J.
Math. Phys. \textbf{42} (2001) 2929, arXiv:hep-th/0001053.

\bibitem{MO2} J. Maldacena, H. Ooguri, and J. Son, {\it Strings in AdS$_3$ and the SL(2,R) WZW Model. Part 2: Euclidean Black Hole}, J. Math. Phys. {\bf 42}(2001) 2961.

\bibitem{MO3} J.M. Maldacena and H. Ooguri, \textit{Strings in AdS}$_{%
\mathit{3}}$\textit{\ and the SL(2,R) WZW Model. Part 3: Correlation
Functions}, Phys. Rev. \textbf{D65} (2002) 106006, arXiv:hep-th/0111180.

\bibitem{Rastelo} G. Giribet, A. Pakman and L. Rastelli, \textit{Spectral
Flow in AdS(3)/CFT(2)}, JHEP \textbf{0806} (2008) 013, arXiv:0712.3046.

\bibitem{CN} C. Cardona and C. N\'{u}\~{n}ez, {\it Three-point functions in superstring theory on AdS(3) x S$^3$ x T$^4$}, JHEP {\bf 0906} (2009) 009.

\bibitem{GN3} G. Giribet and C. N\'{u}\~{n}ez, \textit{Correlators in AdS}$%
_{3}$\textit{\ string theory}, JHEP \textbf{0106} (2001) 010,
arXiv:hep-th/0105200.

\bibitem{GN2} G. Giribet and C. N\'{u}\~{n}ez, \textit{Aspects of the free
field description of string theory on AdS}$_{3}$, , JHEP \textbf{0006}
(2000) 033, arXiv:hep-th/0006070.

\bibitem{succint} G. Giribet, {\it On spectral flow symmetry and Knizhnik-Zamolodchikov equation}, Phys.Lett. {\bf B628} (2005) 148.

\bibitem{R} S. Ribault, \textit{Knizhnik-Zamolodchikov equations and
spectral flow in AdS3 string theory}, JHEP \textbf{0509} (2005) 045,
arXiv:hep-th/0507114.

\bibitem{GKS} A. Giveon, D. Kutasov, and N. Seiberg, {\it Comments on string theory on AdS(3)}, Adv. Theor. Math. Phys. {\bf 2} (1998) 733.

\bibitem{Satoh1} N. Ishibashi, K. Okuyama, and Y. Satoh, {\it Path Integral Approach to String Theory on AdS$_3$}, Nucl. Phys. {\bf B588} (2000) 149.

\bibitem{Satoh2} K. Hosomichi, K. Okuyama, and Y. Satoh, {\it Free Field Appoach to String Theory on AdS$_3$}, Nucl. Phys. {\bf B598} (2001) 451.

\bibitem{Gawedzky} K. Gawedzki, {\it Noncompact WZW conformal field theories}, Proceedings of ''New symmetry principles in quantum field theory´´, Cargese 1991, 247.

\bibitem{Wakimoto} M. Wakimoto, \textit{Fock representations of the affine
Lie algebra A(1)}$^{1}$, Comm. Math. Phys. \textbf{111} (1986) 75.

\bibitem{FZZ} V. Fateev, A. Zamolodchikov, and Al. Zamolodchikov,
unpublished.

\bibitem{FH} T. Fukuda and K. Hosomichi, {\it Three point functions in sine-Liouville theory}, JHEP {\bf 0109} (2001) 003.

\bibitem{Gaston} G. Giribet, \textit{Violating the string winding number
maximally in Anti-de Sitter space}, Phys. Rev. \textbf{D84} (2011) 024045.

\bibitem{Witten} E. Witten, {\it String theory and black holes}, Phys. Rev. {\bf D44} (1991) 314.

\bibitem{BK} M. Bershadsky and D. Kutasov, \textit{Comment of gauged WZW
theory}, Phys. Lett. \textbf{B266} (1991) 345.

\bibitem{DVV} R. Dijkgraaf, H. Verlinde, and E. Verlinde, {\it String propagation in a black hole geometry}, Nucl. Phys. {\bf B371} (1992).

\bibitem{Becker} K. Becker and M. Becker, \textit{Interactions in the
SL(2,R)/U(1) Black Hole Background}, Nucl. Phys. \textbf{B418} (1994) 206,
arXiv:hep-th/9310046.

\bibitem{Dotsenko1} V. Dotsenko, \textit{Solving the SU(2) conformal field
theory using the Wakimoto free field representation}, Nucl. Phys. \textbf{B
338} (1990) 747.

\bibitem{Dotsenko2} V. Dotsenko, \textit{The free field representation of
the su(2) conformal field theory}, Nucl. Phys. \textbf{B358} (1991) 547.

\bibitem{Andreev} O. Andreev, {\it On Affine Lie Superalgebras, AdS$_3$/CFT Correspondence And World-Sheets For World-Sheets}, Nucl. Phys. {\bf B552} (1999) 169.

\bibitem{GoulianLi} M. Goulian and M. Li, {\it Correlation functions in Liouville theory}, Phys. Rev. Lett. {\bf 66} (1991) 2051.

\bibitem{DF1} V. Dotsenko and V. Fateev, {\it Operator Algebra of Two-Dimensional Conformal Theories with Central Charge $c \leq 1$}, Phys. Lett. {\bf B154} (1985) 291.

\bibitem{DF2} V. Dotsenko and V. Fateev, {\it Four Point Correlation Functions and the Operator Algebra in the Two-Dimensional Conformal Invariant Theories with the Central Charge $c < 1$}, Nucl. Phys. {\bf B251} (1985) 691.

\bibitem{DF3} V. Dotsenko and V. Fateev, {\it Conformal Algebra and Multipoint Correlation Functions in Two-Dimensional Statistical Models}, Nucl. Phys. {\bf B240} (1984) 312.

\bibitem{RibaultTeschner} S. Ribault and J. Teschner, \textit{H(3)+
correlators from Liouville theory}, JHEP \textbf{0506} (2005) 014,
arXiv:hep-th/0502048.

\bibitem{Stoyanovsky} A.V. Stoyanovsky, \textit{A relation between the
Knizhnik--Zamolodchikov and Belavin--Polyakov--Zamolodchikov systems of
partial differential equations, }arXiv:math-ph/0012013.

\bibitem{HikidaSchomerus} Y. Hikida and V. Schomerus, \textit{H}$_{3}^{+}$%
\textit{\ WZNW model from Liouville field theory}, JHEP \textbf{0710} (2007)
064, arXiv:0706.1030.

\bibitem{Ooguri} J. de Boer, H. Ooguri, H. Robins, and J. Tannenhauser, {\it String theory on AdS(3)}, JHEP {\bf 9812} (1998) 026.

\bibitem{Teschner1} J. Teschner, {\it Operator product expansion and factorization in the H+(3) WZNW model}, Nucl. Phys. {\bf B571} (2000) 555.

\bibitem{Teschner2} J. Teschner, {\it On structure constants and fusion rules in the SL(2,C)/SU(2) WZNW model}, Nucl. Phys. {\bf B546} (1999) 390.

\bibitem{Yo} G. Giribet, {\it The String Theory on AdS$_3$ as a Marginal Deformation of a Linear Dilaton Background}, Nucl. Phys. {\bf B737} (2006) 209.

\bibitem{HM} D. Hofman and C. N\'{u}\~{n}ez, {\it Free field realization of superstring theory on AdS(3)}, JHEP {\bf 0407} (2004) 019. 

\bibitem{HikidaSchomerus2} Y. Hikida and V. Schomerus, \textit{H}$_{3}^{+}$%
\textit{ The FZZ-Duality Conjecture - A Proof}, JHEP \textbf{0903} (2009)
095.

\bibitem{KKK} V. Kazakov, I. Kostov and D. Kutasov, {\it A Matrix Model for the Two Dimensional Black Hole}, Nucl.Phys. {\bf B622} (2002) 141.


\end{thebibliography}
\end{document}